\documentclass[12pt,english]{article}
\usepackage[T1]{fontenc}
\usepackage[latin1]{inputenc}
\usepackage{a4wide}
\usepackage{amsmath}
\usepackage{graphicx}
\usepackage{amssymb}

\makeatletter

\newcommand{\lyxline}[1]{
  {#1 \vspace{1ex} \hrule width \columnwidth \vspace{1ex}}
}

 \newcommand{\lyxaddress}[1]{
   \par {\raggedright #1
   \vspace{1.4em}
   \noindent\par}
 }
 \usepackage{verbatim}
 \newenvironment{lyxlist}[1]
   {\begin{list}{}
     {\settowidth{\labelwidth}{#1}
      \setlength{\leftmargin}{\labelwidth}
      \addtolength{\leftmargin}{\labelsep}
      }}
   {\end{list}}

\newcommand{\e}{\mbox{e}}
\def\slash#1{\setbox0=\hbox{$#1$}  
   \dimen0=\wd0     
   \setbox1=\hbox{/} \dimen1=\wd1  
   \ifdim\dimen0>\dimen1   
      \rlap{\hbox to \dimen0{\hfil/\hfil}} 
      #1     
   \else     
      \rlap{\hbox to \dimen1{\hfil$#1$\hfil}} 
      /      
   \fi}      %

\renewcommand{\i}{\mathrm{i}}
\renewcommand{\d}{\mathrm{d}}

\newcommand{\M}{\mathcal{M}}

\usepackage{babel}
\makeatother
\begin{document}

\title{\textbf{Soft Pion Emission in}\\
\textbf{Hard Exclusive Pion Production}}

\author{Maxim V.~Polyakov$^{a,b}$, Simone Stratmann$^{b}$}

\maketitle

\lyxaddress{$^{a}$ \emph{St.~Petersburg Nuclear Physics Institute, Gatchina,
St.~Petersburg, Russia}\\
\emph{$^{b}$ Institut für Theoretische Physik II, Ruhr-Universität,
D-44780 Bochum, Germany}}

\begin{abstract}
We investigate hard exclusive reactions on the nucleon with soft pion
emission. A parametrization of corresponding hadronic matrix elements
in terms of parton distributions for final pion-nucleon state is provided.
These distributions are calculated in terms of nucleon and pion GPDs
and the pion distribution amplitude via soft-pion theorems. Some observables
for the process of hard charged pion production on the proton with
soft pion emission are computed.
\end{abstract}

\section{Motivation and outline}

The field of hard exclusive reactions has been studied intensively
during the past decade. The investigations have proceeded in the theoretical
as well as in the experimental sector; reviews are given e.g.~in
references \cite{Goeke:2001tz,Ji:1998pc,Radyushkin:2000uy,Diehl:2003ny,Belitsky:2005qn}.
Two prominent representatives of the hard exclusive processes are
deeply virtual Compton scattering (DVCS) and hard meson production
(HMP). The nucleon properties which enter these reactions are formulated
in terms of generalized parton distributions (GPDs). On the one hand,
these functions can be viewed as generalizations of the usual forward
parton distributions, on the other hand they are directly related
to nucleon form factors through their moments. So it has been argued
that hard exclusive reactions can provide useful new insights into
the partonic nucleon structure which are not accessible through the
usual electroweak probes. In this context, for example, the form factors
of the energy-momentum tensor have been discussed, see e.g.~references
\cite{Ji:1997nm,Polyakov:2002yz}. Additionally, in the case of meson
production, one can obtain information about the involved distributions
amplitudes.

In this article, we investigate the situation when in a hard exclusive
reaction instead of the final nucleon a nucleon-pion state with low
invariant mass appears. Since it is produced close to the threshold,
the pion is denoted as \emph{soft}. On the experimental side, a separation
of a particular hard reaction with and without soft pion emission
cannot always be guaranteed. In this sense, the process with soft
pion can be viewed as a contamination of the fully exclusive DVCS
or HMP, and theoretical estimates about this disturbing background
are desirable.

Apart from such practical considerations, these new reactions are
worth being studied in their own right. They provide an opportunity
to investigate soft pion emission from the nucleon induced by nonlocal
lightcone operators as opposed to the local vector or axial operators
to which we are restricted in usual electroweak pion production. Therefore,
analogously to how a soft process such as pion-electroproduction can
provide information about nucleon form factors, hard processes with
soft pion emission might contribute to a better understanding of quantities
such as generalized parton distributions.

Guichon et.~al.~have addressed this question for the process of
DVCS with soft pion production \cite{Guichon:2003ah}. For the calculation
of the pertinent hadronic matrix element, they presented a soft-pion
theorem based on current algebra and chiral symmetry. Moreover, they
modeled the effect of the $\Delta(1232)$ re\-so\-nance which is
located not far from the pion production threshold. By this method,
they could give predictions for certain cross sections and asymmetries.
However, in all their considerations, the region of small momentum
transfer was explicitely excluded.

In the following, we present our approach of calculating soft pion
emission in hard exclusive reactions. Based on polology arguments,
{}``PCAC'', and certain properties of the chiral symmetry transformation,
we derive corresponding soft-pion theorems. These differ from the
results in \cite{Guichon:2003ah} through additional pion pole terms.
The implementation of these new contributions allows in particular
to extend the region of applicability down to small momentum transfer.
Using this improved result, we calculate the effect of soft pions
in hard $\pi^{+}$ production off the proton.

The outline is as follows. Basic kinematical considerations are given
in section 2. In section 3, we provide a parametrization of the matrix
elements for pion emission induced by twist-2 lightcone operators.
We denote the invariant functions that come up in this procedure as
pion-nucleon ($\pi N$) parton distributions. These functions are
the generalizations of GPDs for the case of pion emission. We discuss
some of their properties, in particular their behavior at the pion
threshold and the meaning of some moments. In section 4, we give a
detailed derivation and the results of soft-pion theorems for several
twist-2 operators. We check that in certain limiting cases our expressions
are consistent with previous calculations. Finally, in section 5,
we apply these results to the process of hard pion production off
the proton with soft pion emission near threshold. The amplitude of
the process for arbitrary isospins is given. Further, we provide the
following numerical estimates for hard $\pi^{+}$ production: the
transverse spin asymmetry of the process $\gamma^{*}+\textrm{p}\rightarrow\pi^{+}+N+\pi_{\mathrm{soft}}$
is calculated, and the contamination of the longitudinal cross section
and the transverse spin asymmetry of the pure process $\gamma^{*}+\textrm{p}\rightarrow\pi^{+}+\textrm{n}$
through soft-pion admixture is determined.

\section{Kinematics}

\label{section_kinematics}

We consider the collision of a virtual photon $\gamma^{*}$ with momentum
$q$ and a nucleon $N$ with momentum $p$ and spin $S$. In the final
state, we have a nucleon with momentum $p'$ and spin $S'$, a pion
$\pi$ with momentum $k$ and isospin $a$, and either a real photon
$\gamma$ (DVCS) or a specified meson $M$ with momentum $q'$: \[
\gamma^{*}(q)+N(p,S)\rightarrow N(p',S')+\pi^{a}(k)+\left\{ \begin{array}{c}
\gamma(q')\\
M(q')\end{array}\right..\]
It is useful to define an average momentum $\bar{p}$ and a momentum
transfer $\Delta$ in the following way:\begin{equation}
\bar{p}\equiv\frac{p+p'+k}{2},\qquad\Delta\equiv q-q'=p'+k-p.\end{equation}
Further, we introduce the Lorentz invariants%
\begin{figure}[!t]
\begin{center}\includegraphics[%
  width=0.40\textwidth]{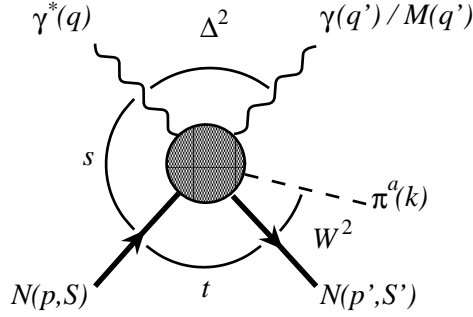}\end{center}

\caption{\label{Invariants.fig}Particle momenta and invariants for the exclusive
reaction with soft pion emission.}\lyxline{\normalsize}

\end{figure}
 \begin{equation}
s\equiv(p+q)^{2},\qquad t\equiv(p-p')^{2},\qquad u\equiv(p-k)^{2},\qquad W^{2}\equiv(p'+k)^{2},\end{equation}
see also the illustration in figure \ref{Invariants.fig} (we point
to the fact that here $\Delta^{2}$ is \emph{not} identical to the
nucleon momentum transfer $t$ unlike in the case without pion). The
kinematical region of hard scattering is characterized by a large
photon virtuality $Q^{2}$, a large energy $\nu$ of the virtual photon
in the target rest frame, and a fixed Bjorken variable $x_{B}$: \begin{equation}
Q^{2}\equiv-q^{2},\; M\nu\equiv p\cdot q\quad\gg\quad-\Delta^{2},\;-t,\; W^{2},\; M^{2},\; m_{\pi}^{2},\; q'^{2},\end{equation}
\begin{equation}
x_{B}\equiv\frac{Q^{2}}{2M\nu}.\end{equation}
Here, $M$ and $m_{\pi}$ denote the nucleon and pion mass, respectively.

We shall refer to the pion $\pi(k)$ as \emph{soft} if it appears
sufficiently close to its production threshold in the following sense.
Directly at the threshold, the variables $W^{2}$ and $u$ are fixed
as\begin{equation}
W_{\mathrm{th}}^{2}=(M+m_{\pi})^{2},\qquad u_{\mathrm{th}}=(M-m_{\pi})^{2}+\frac{m_{\pi}}{M}t.\label{W_u_th}\end{equation}
We allow for deviations from these values that are of the size\begin{equation}
W^{2}-W_{\mathrm{th}}^{2}\lesssim Mm_{\pi},\qquad-(u-u_{\mathrm{th}})\lesssim Mm_{\pi}.\end{equation}
Moreover, we require for the nucleon momentum transfer that\begin{equation}
-t\lesssim M^{2}.\label{range_t}\end{equation}
Relations (\ref{W_u_th}) to (\ref{range_t}) settle the soft pion
kinematics (note that they imply $-\Delta^{2}\lesssim M^{2}$). Roughly
speaking, we can summarize these conditions simply as\begin{equation}
k=\mathcal{O}(m_{\pi}),\qquad p,p'=\mathcal{O}(M).\end{equation}

Let us now turn to a Sudakov decomposition of the particle momenta.
Two lightcone vectors $\tilde{n}$ and $n$ are defined in a frame
where $\bar{p}=(\bar{p}_{0},0,0,\bar{p}_{z})$ and $q=(q_{0},0,0,q_{z})$
via\begin{equation}
\tilde{n}=(\bar{p}_{0}+\bar{p}_{z})(1,0,0,1),\qquad n=\frac{1}{2(\bar{p}_{0}+\bar{p}_{z})}(1,0,0,-1).\end{equation}
Consequently, we obtain the average momentum and the virtual photon
momentum in the form\begin{equation}
\bar{p}=\tilde{n}+\frac{\bar{M}^{2}}{2}n,\qquad q=-2\xi\tilde{n}+\frac{Q^{2}}{4\xi}n,\end{equation}
with\begin{equation}
\bar{M}^{2}=\bar{p}^{2}=\frac{M^{2}+W^{2}}{2}-\frac{\Delta^{2}}{4},\qquad-2\xi=\frac{\bar{p}\cdot q}{\bar{M}^{2}}-\sqrt{\left(\frac{\bar{p}\cdot q}{\bar{M}^{2}}\right)^{2}+\frac{Q^{2}}{\bar{M}^{2}}}.\end{equation}
The decompositions of some other vectors to leading order in $1/Q^{2}$
read\begin{eqnarray}
\Delta & = & -2\xi\tilde{n}+\left(\bar{M}^{2}\xi+\frac{W^{2}-M^{2}}{2}\right)n+\Delta_{\bot},\\
p & = & (1+\xi)\tilde{n}+\left[\frac{\bar{M}^{2}}{2}(1-\xi)-\frac{W^{2}-M^{2}}{4}\right]n-\frac{\Delta_{\bot}}{2},\\
p'+k & = & (1-\xi)\tilde{n}+\left[\frac{\bar{M}^{2}}{2}(1+\xi)+\frac{W^{2}-M^{2}}{4}\right]n+\frac{\Delta_{\bot}}{2}.\end{eqnarray}
Note that these expressions lead immediately to the relations\begin{equation}
x_{B}=\frac{2\xi}{1+\xi}\end{equation}
and\begin{equation}
\Delta_{\bot}^{2}=-(1-\xi^{2})(|\Delta^{2}|-|\Delta^{2}|_{\mathrm{min}}),\end{equation}
with the minimal value of the momentum transfer given by\begin{equation}
|\Delta^{2}|_{\mathrm{min}}=\frac{2\xi}{1-\xi^{2}}[(W^{2}+M^{2})\xi+W^{2}-M^{2}]=\frac{x_{B}^{2}M^{2}+x_{B}(W^{2}-M^{2})}{1-x_{B}}.\end{equation}
Setting $k=0$ and $W=M$ everywhere, one recovers the well-known
formulas of the familiar case without pion.

Finally, in order to quantify how the final nucleon and pion share
the momentum with respect to the lightcone direction $\tilde{n}$,
we further introduce a variable $\alpha$ such that\begin{equation}
k\cdot n=\alpha(1-\xi),\qquad p'\cdot n=(1-\alpha)(1-\xi).\end{equation}

\section{Formulation of pion-nucleon parton distributions}

\subsection{General parametrization}

In the case of ordinary DVCS or hard meson production, the factorization
of the amplitude leads to nucleon matrix elements of lightcone operators.
These nonperturbative objects are parametrized in terms of generalized
parton distributions. For hard meson production, additionally the
distribution amplitude of the final meson appears.

In the present situation, where we take into account an additional
pion, we arrive at matrix elements of lightcone operators with initial
nucleon and final pion-nucleon state. In dealing with these objects,
our first step is to provide a parametrization. Let us start with
the isoscalar quark operator of vector type:\begin{equation}
\int\frac{\d\lambda}{2\pi}\e^{\i\lambda x\bar{p}\cdot n}\langle N(p')\pi^{a}(k)|\bar{\psi}(-\lambda n/2)\slash{n}\psi(\lambda n/2)|N(p)\rangle=\frac{\i g_{A}}{Mf_{\pi}}\sum_{i=1}^{4}\bar{U}(p')\,\Gamma_{i}H_{i}^{(0)}\tau^{a}U(p)\label{Hi0_def}\end{equation}
where we have introduced the Dirac matrices\begin{equation}
\Gamma_{1}=\gamma_{5},\qquad\Gamma_{2}=\frac{M\slash{n}}{n\cdot\bar{p}}\gamma_{5},\qquad\Gamma_{3}=\frac{\slash{k}}{M}\gamma_{5},\qquad\Gamma_{4}=\frac{\slash{k}\slash{n}}{n\cdot\bar{p}}\gamma_{5}\label{Gamma_i}\end{equation}
with the definitions $\Delta=p'+k-p$ and $\bar{p}=(p'+k+p)/2$. The
constant $g_{A}$ is the isovector axial coupling constant, and $f_{\pi}$
is the pion decay constant normalized according to the experimental
value of $93\textrm{ MeV}$. Including these constants into the definition
will be convenient in later calculations. Here and in the following,
the insertion of the appropriate gauge links within the operators
is always understood. The functions $H_{i}^{(0)}$, that we shall
call pion-nucleon parton distributions, depend on the momentum fraction
$x$ as well as on five further quantities that can be built from
the vectors $n$, $p$, $p'$, and $k$ which we have at our disposal.
We choose \begin{equation}
H_{i}^{(0)}=H_{i}^{(0)}(x,\xi,\Delta^{2},\alpha,t,W^{2}),\end{equation}
where in the present context of parametrization the definitions of
the variables are\begin{equation}
\xi=-\frac{n\cdot\Delta}{2n\cdot\bar{p}},\qquad\alpha=\frac{n\cdot k}{n\cdot(p'+k)},\qquad t=(p'-p)^{2},\qquad W^{2}=(p'+k)^{2}.\end{equation}
In a similar way, we introduce the isovector even and odd distributions
$H_{i}^{(+)}$ and $H_{i}^{(-)}$,\begin{eqnarray}
\lefteqn{\int\frac{\d\lambda}{2\pi}\e^{\i\lambda x\bar{p}\cdot n}\langle N'\pi^{a}|\bar{\psi}(-\lambda n/2)\slash{n}\tau^{b}\psi(\lambda n/2)|N\rangle}\nonumber \\
 &  & =\frac{\i g_{A}}{Mf_{\pi}}\sum_{i=1}^{4}\bar{U}'\Gamma_{i}(\delta^{ab}H_{i}^{(+)}+\i\varepsilon^{abc}\tau^{c}H_{i}^{(-)})U,\label{Hipm_def}\end{eqnarray}
which depend on the same set of arguments as $H^{(0)}$, of course.

The matrix elements of the quark operators of axial type differ from
the previous ones only through the insertion of a matrix $\gamma_{5}$.
Therefore, we can define in an analogous manner isoscalar distributions,\begin{equation}
\int\frac{\d\lambda}{2\pi}\e^{\i\lambda x\bar{p}\cdot n}\langle N'\pi^{a}|\bar{\psi}(-\lambda n/2)\slash{n}\gamma_{5}\psi(\lambda n/2)|N\rangle=\frac{\i g_{A}}{Mf_{\pi}}\sum_{i=1}^{4}\bar{U}'\Gamma_{i}\gamma_{5}\tau^{a}\tilde{H}_{i}^{(0)}U\label{HiTilde0_def}\end{equation}
and isovector distributions,\begin{eqnarray}
\lefteqn{\int\frac{\d\lambda}{2\pi}\e^{\i\lambda x\bar{p}\cdot n}\langle N'\pi^{a}|\bar{\psi}(-\lambda n/2)\slash{n}\gamma_{5}\tau^{b}\psi(\lambda n/2)|N\rangle}\nonumber \\
 &  & =\frac{\i g_{A}}{Mf_{\pi}}\sum_{i=1}^{4}\bar{U}'\Gamma_{i}\gamma_{5}(\delta^{ab}\tilde{H}_{i}^{(+)}+\i\varepsilon^{abc}\tau^{c}\tilde{H}_{i}^{(-)})U.\label{HiTildepm_def}\end{eqnarray}
 Furthermore, the gluon matrix element can be parametrized as\begin{equation}
\int\frac{\d\lambda}{2\pi}\e^{\i\lambda x\bar{p}\cdot n}\langle N'\pi^{a}|F^{\mu\rho}(-\lambda n/2){F_{\rho}}^{\nu}(\lambda n/2)n_{\mu}n_{\nu}|N\rangle=\frac{\i g_{A}}{Mf_{\pi}}\frac{n\cdot\bar{p}}{2}\sum_{i=1}^{4}\bar{U}'\Gamma_{i}\tau^{a}xH_{i}^{(G)}U.\label{Hig_def}\end{equation}

At this point we have to comment on a corresponding parametrization
that was previously given by Bl\"umlein et.~al. \cite{Blumlein:2001sb}.
The authors came to the conclusion that \emph{five} functions are
necessary for a complete description. However, as demonstrated explicitely
in appendix E of reference \cite{Str03}, one can show that one of
their functions can actually be reexpressed in terms of the others,
i.e.~the structures given in their parametrization are not linearly
independent if one takes into account the Dirac equation.

\subsection{Reduction at the pion threshold}

Within the soft pion kinematics described in section \ref{section_kinematics},
all variables are required to be close to their values at the pion
production threshold. Therefore, the \emph{exact} threshold kinematics
can serve as a reasonable approximation e.g.~for the calculation
of scattering amplitudes. For this purpose, let us consider the threshold
case and provide some useful relations.

First, we recall that the invariant mass of the final nucleon-pion
system is equal to\begin{equation}
W_{\mathrm{th}}^{2}=(M+m_{\pi})^{2}.\end{equation}
This is equivalent to the statement that the four-momenta $k$ and
$p'$ are proportional,\begin{equation}
k=\frac{m_{\pi}}{M}p',\label{proportional}\end{equation}
hence the number of independent invariants is reduced. In particular,
we find that $\Delta^{2}$ can be expressed through $t$ and that
$\alpha$ becomes a constant:\begin{equation}
\Delta_{\mathrm{th}}^{2}=\frac{M+m_{\pi}}{M}t+m_{\pi}^{2},\qquad\alpha_{\mathrm{th}}=\frac{m_{\pi}}{M+m_{\pi}}.\label{momentum_transfer_th}\end{equation}

As a further consequence of the relation (\ref{proportional}), the
Dirac structures involving e.g.~$\Gamma_{1}=\gamma_{5}$ and $\Gamma_{3}=\slash k\gamma_{5}/M$
are no longer linearly independent. Therefore, the number of $\pi N$
distributions can be reduced by two. For example, in the case of the
vector distributions $H_{i}$, we arrive at\begin{equation}
\int\frac{\d\lambda}{2\pi}\e^{\i\lambda x\bar{p}\cdot n}\langle N'\pi^{a}|\bar{\psi}(-\lambda n/2)\slash{n}\psi(\lambda n/2)|N\rangle=\frac{\i g_{A}}{Mf_{\pi}}\bar{U}'\left[H_{1\mathrm{th}}^{(0)}+\frac{M\slash{n}}{n\cdot\bar{p}}H_{2\mathrm{th}}^{(0)}\right]\gamma_{5}\tau^{a}U\label{NNpiGPDth}\end{equation}
and\begin{eqnarray}
\lefteqn{\int\frac{\d\lambda}{2\pi}\e^{\i\lambda x\bar{p}\cdot n}\langle N'\pi^{a}|\bar{\psi}(-\lambda n/2)\slash{n}\tau^{b}\psi(\lambda n/2)|N\rangle}\nonumber \\
 &  & =\frac{\i g_{A}}{Mf_{\pi}}\bar{U}'\left[(\delta^{ab}H_{1\mathrm{th}}^{(+)}+\i\varepsilon^{abc}\tau^{c}H_{1\mathrm{th}}^{(-)})+\frac{M\slash{n}}{n\cdot\bar{p}}(\delta^{ab}H_{2\mathrm{th}}^{(+)}+\i\varepsilon^{abc}\tau^{c}H_{2\mathrm{th}}^{(-)})\right]\gamma_{5}U,\end{eqnarray}
with the threshold pion-nucleon distributions $H_{1\mathrm{th}}$
and $H_{2\mathrm{th}}$ given by\begin{eqnarray}
H_{1\mathrm{th}}^{(0,\pm)}(x,\xi,t) & = & H_{1}^{(0,\pm)}(x,\xi,\Delta_{\mathrm{th}}^{2},\alpha_{\mathrm{th}},t,W_{\mathrm{th}}^{2})+\frac{m_{\pi}}{M}H_{3}^{(0,\pm)}(x,\xi,\Delta_{\mathrm{th}}^{2},\alpha_{\mathrm{th}},t,W_{\mathrm{th}}^{2})\label{H1th}\\
H_{2\mathrm{th}}^{(0,\pm)}(x,\xi,t) & = & H_{2}^{(0,\pm)}(x,\xi,\Delta_{\mathrm{th}}^{2},\alpha_{\mathrm{th}},t,W_{\mathrm{th}}^{2})+\frac{m_{\pi}}{M}H_{4}^{(0,\pm)}(x,\xi,\Delta_{\mathrm{th}}^{2},\alpha_{\mathrm{th}},t,W_{\mathrm{th}}^{2}).\label{H2th}\end{eqnarray}
Analogously, we define $\tilde{H}_{1\mathrm{th}}^{(0,\pm)}$ and $\tilde{H}_{2\mathrm{th}}^{(0,\pm)}$
for the axial operators and $H_{1\mathrm{th}}^{(G)}$ and $H_{2\mathrm{th}}^{(G)}$
for the gluon operator.

\subsection{Moments of the $\pi N$ distributions}

\label{section_moments}

The moments of ordinary nucleon GPDs are polynomials in the skewedness
variable, where the coefficients give the nucleon form factors of
the corresponding local twist-2 operators \cite{Ji:1997nm,Ji:1997gm}.
For example, taking the first moments of the quark GPDs yields\begin{equation}
\int\d x\, H(x,\xi,\Delta^{2})=F_{1}(\Delta^{2}),\qquad\int\d x\, E(x,\xi,\Delta^{2})=F_{2}(\Delta^{2}),\end{equation}
\begin{equation}
\int\d x\,\tilde{H}(x,\xi,\Delta^{2})=G_{A}(\Delta^{2}),\qquad\int\d x\,\tilde{E}(x,\xi,\Delta^{2})=G_{P}(\Delta^{2}),\end{equation}
where $F_{1}$ and $F_{2}$ are the Dirac and Pauli form factor and
$G_{A}$ and $G_{P}$ the axial and pseudoscalar form factor, respectively.
Concerning the second moment, we have e.g.\begin{eqnarray}
\int\d x\, x\left[H^{(S)}(x,\xi,\Delta^{2})+\frac{1}{2}H^{(G)}(x,\xi,\Delta^{2})\right] & = & A(\Delta^{2})+C(\Delta^{2})\,(2\xi)^{2},\\
\int\d x\, x\left[E^{(S)}(x,\xi,\Delta^{2})+\frac{1}{2}E^{(G)}(x,\xi,\Delta^{2})\right] & = & B(\Delta^{2})-C(\Delta^{2})\,(2\xi)^{2},\end{eqnarray}
where $A$, $B$, and $C$ are the form factors of the energy-momentum
tensor. Similar polynomiality conditions hold for the moments of the
$\pi N$ distributions, their moments are polynomials in the variables
$\xi$ and $\alpha$. We shall demonstrate this now explicitely for
two examples.

\subsubsection{First moment: The local limit}

The first moments of the $\pi N$ distributions $H_{i}^{(0,\pm)}$
are related to the form factors of the matrix element which describes
pion emission from the nucleon induced by the the local vector current
(the hadronic ingredient of the pion electroproduction amplitude).
For our purposes, the following parametrization in terms of pion emission
form factors $A_{i}$ is convenient: \begin{eqnarray}
\lefteqn{\langle N(p')\,\pi^{a}(k)|\bar{\psi}\gamma^{\mu}\left\{ \begin{array}{c}
1\\
\tau^{b}\end{array}\right\} \psi|N(p)\rangle}\nonumber \\
 & = & \frac{\i g_{A}}{Mf_{\pi}}\sum_{i=1}^{8}\bar{U}(p')\,\left\{ \begin{array}{c}
\tau^{a}A_{i}^{(0)}\\
\delta^{ab}A^{(+)}+\i\varepsilon^{abc}\tau^{c}A^{(-)}\end{array}\right\} \Gamma_{i}^{\mu}U(p),\label{electroproduction_amplitude}\end{eqnarray}
where the set of Dirac matrices is chosen as\begin{equation}
\{\Gamma_{1}^{\mu},\ldots,\Gamma_{8}^{\mu}\}=\{\bar{p}^{\mu},\,\Delta^{\mu},\, k^{\mu},\,\gamma^{\mu},\,\slash k\bar{p}^{\mu},\,\slash k\Delta^{\mu},\,\slash kk^{\mu},\,\slash k\gamma^{\mu}\}\gamma_{5}.\end{equation}
(For a traditional parametrization we refer to Amaldi et.~al.~\cite{Ama79}.)
The form factors are functions of three independent invariants, e.g.~$\Delta^{2}$,
$W^{2}$, and $t$. Current conservation reduces the number of independent
form factors to six:\begin{eqnarray}
(W^{2}-M^{2})A_{1}+2\Delta^{2}A_{2}+(W^{2}+u-2M^{2})A_{3}+4MA_{4}+2(W^{2}-M^{2})A_{8} & = & 0\label{current_conservation1}\\
2A_{4}+(W^{2}-M^{2})A_{5}+2\Delta^{2}A_{6}+2(W^{2}+u-2M^{2})A_{7} & = & 0.\label{current_conservation2}\end{eqnarray}
From the contraction of the matrix element (\ref{electroproduction_amplitude})
with the lightcone vector $n$, it follows that the first moments
of the $\pi N$ distributions are polynomials in $\Delta\cdot n/\bar{p}\cdot n=-2\xi$
and $k\cdot n/\bar{p}\cdot n=\alpha(1-\xi)\equiv\bar{\alpha}$: \begin{equation}
\int\limits _{-1}^{1}\d x\, H_{1}=A_{1}-2\xi A_{2}+\bar{\alpha}A_{3},\qquad M\int\limits _{-1}^{1}\d x\, H_{2}=A_{4},\end{equation}
\begin{equation}
\frac{1}{M}\int\limits _{-1}^{1}\d x\, H_{3}=A_{5}-2\xi A_{6}+\bar{\alpha}A_{7},\qquad\int\limits _{-1}^{1}\d x\, H_{4}=A_{8}.\end{equation}
Note that the current conservation relations (\ref{current_conservation1})
and (\ref{current_conservation2}) impose nontrivial conditions on
the moments of the $\pi N$ distributions.

\subsubsection{Second moment: The energy-momentum tensor}

The second moment of the $\pi N$ distributions $H_{i}^{(0)}$ and
$H_{i}^{(G)}$ is related to the form factors of the amplitude for
pion emission induced by the energy-momentum tensor, which reads\begin{equation}
\mathcal{T}^{\mu\nu}=\frac{\i}{2}\bar{\psi}\gamma^{\{\mu}(\overrightarrow{D}-\overleftarrow{D})^{\nu\}}\psi+\frac{g^{\mu\nu}}{4}F^{\rho\sigma}F_{\rho\sigma}+F^{\mu\rho}{F_{\rho}}^{\nu},\end{equation}
where the curly brackets denote symmetrization of the indices and
$D$ the covariant derivative. We parametrize this amplitude as follows:\begin{equation}
\langle N(p')\,\pi^{a}(k)|\mathcal{T}^{\mu\nu}|N_{i}(p)\rangle=\frac{\i g_{A}}{Mf_{\pi}}\sum_{i=1}^{20}\bar{U}(p')\,\tau^{a}\Gamma_{i}^{\mu\nu}B_{i}U,\end{equation}
where the Dirac matrices are\begin{eqnarray}
\{\Gamma_{1}^{\mu\nu},\ldots,\Gamma_{20}^{\mu\nu}\} & = & \{ g^{\mu\nu},\,\bar{p}^{\mu}\bar{p}^{\nu},\,\Delta^{\mu}\Delta^{\nu},\, k^{\mu}k^{\nu},\,\bar{p}^{\{\mu}\Delta^{\nu\}},\,\bar{p}^{\{\mu}k^{\nu\}},\,\Delta^{\{\mu}k^{\nu\}},\nonumber \\
 &  & \phantom{\{}g^{\mu\nu}\slash k,\,\bar{p}^{\mu}\bar{p}^{\nu}\slash k,\,\Delta^{\mu}\Delta^{\nu}\slash k,\, k^{\mu}k^{\nu}\slash k,\,\bar{p}^{\{\mu}\Delta^{\nu\}}\slash k,\,\bar{p}^{\{\mu}k^{\nu\}}\slash k,\,\Delta^{\{\mu}k^{\nu\}}\slash k,\nonumber \\
 &  & \phantom{\{}\gamma^{\{\mu}\bar{p}^{\nu\}},\gamma^{\{\mu}\,\Delta^{\nu\}},\gamma^{\{\mu}k^{\nu\}},\nonumber \\
 &  & \phantom{\{}\slash k\gamma^{\{\mu},\,\slash k\gamma^{\{\mu}\bar{p}^{\nu\}},\,\slash k\gamma^{\{\mu}\Delta^{\nu\}},\,\slash k\gamma^{\{\mu}k^{\nu\}}\}\gamma_{5}.\end{eqnarray}
As in the case of the vector current, the form factors $B_{i}$ are
functions of e.g.~$\Delta^{2}$, $W^{2}$, and $t$. From energy-momentum
conservation, we have derived the following set of constraints:\begin{eqnarray}
\lefteqn{4B_{1}+4\Delta^{2}B_{3}+(W^{2}-M^{2})B_{5}}\nonumber \\
 &  & +(W^{2}+u-2M^{2})B_{7}+4MB_{16}+2(W^{2}-M^{2})B_{19}=0,\label{T_current_conservation1}\end{eqnarray}
\begin{equation}
2(W^{2}-M^{2})B_{2}+2\Delta^{2}B_{5}+(W^{2}+u-2M^{2})B_{6}+4MB_{15}+2(W^{2}-M^{2})B_{18}=0,\end{equation}
\begin{equation}
4(W^{2}+u-2M^{2})B_{4}+(W^{2}-M^{2})B_{6}+2\Delta^{2}B_{7}+4MB_{17}+2(W^{2}-M^{2})B_{20}=0,\end{equation}
\begin{equation}
4B_{8}+4\Delta^{2}B_{10}+(W^{2}-M^{2})B_{12}+(W^{2}+u-2M^{2})B_{14}+2B_{16}=0,\label{T_current_conservation_4}\end{equation}
\begin{equation}
2(W^{2}-M^{2})B_{9}+2\Delta^{2}B_{12}+(W^{2}+u-2M^{2})B_{13}+2B_{15}=0,\end{equation}
\begin{equation}
2(W^{2}+u-2M^{2})B_{11}+(W^{2}-M^{2})B_{13}+2\Delta^{2}B_{14}+2B_{17}=0,\end{equation}
\begin{equation}
(W^{2}-M^{2})B_{15}+2\Delta^{2}B_{16}+(W^{2}+u-2M^{2})B_{17}=0,\end{equation}
\begin{equation}
(W^{2}-M^{2})B_{18}+2\Delta^{2}B_{19}+(W^{2}+u-2M^{2})B_{20}=0,\end{equation}
which reduces the number of independent functions $B_{i}$ to twelve.
The polynomiality conditions read\begin{equation}
\int\limits _{-1}^{1}\d x\, x\left(H_{1}^{(0)}+\frac{1}{2}H_{1}^{(G)}\right)=B_{2}+B_{3}(2\xi)^{2}+B_{4}\bar{\alpha}^{2}+B_{5}(-2\xi)+B_{6}\bar{\alpha}+B_{7}(-2\xi\bar{\alpha}),\end{equation}
\begin{equation}
M\int\limits _{-1}^{1}\d x\, x\left(H_{2}^{(0)}+\frac{1}{2}H_{2}^{(G)}\right)=B_{15}+B_{16}(-2\xi)+B_{17}\bar{\alpha},\end{equation}
\begin{equation}
\frac{1}{M}\int\limits _{-1}^{1}\d x\, x\left(H_{3}^{(0)}+\frac{1}{2}H_{3}^{(G)}\right)=B_{9}+B_{10}(2\xi)^{2}+B_{11}\bar{\alpha}^{2}+B_{12}(-2\xi)+B_{13}\bar{\alpha}+B_{14}(-2\xi\alpha),\end{equation}
\begin{eqnarray}
\int\limits _{-1}^{1}\d x\, x\left(H_{4}^{(0)}+\frac{1}{2}H_{4}^{(G)}\right) & = & B_{18}+B_{19}(-2\xi)+B_{20}\bar{\alpha}.\end{eqnarray}
Thus we see that from current conservation and polynomiality it possible
to uniquely determine the form factors $B_{i}$ from the second moments
of the $\pi N$ distributions $H_{i}^{(0)}+H_{i}^{(G)}/2$.

\section{Soft-pion theorems for pion emission from the nucleon induced by
lightcone operators}

In the last section, we have presented a parametrization of matrix
elements for pion emission from the nucleon which is induced by
twist-2 quark or gluon lightcone operators. Now, we turn to the
calculation of these objects in the soft-pion region. For this, we
rely on po\-lo\-lo\-gy arguments, PCAC, and current algebra. The
basic ideas are similar to those in the work of Guichon
et.~al.~\cite{Guichon:2003ah}. However, while Guichon
et.~al.~excluded small momentum transfer, we shall consider this
region in our derivation as well. Actually, for certain operators
it was already shown in reference \cite{KPS} that at small
momentum transfer, the soft-pion theorems match the tree level
results of a chiral perturbation theory treatment. This refutes opposite
claim of ref.~\cite{Chen:2003jm}. The results of \cite{Guichon:2003ah}
and \cite{KPS} were confirmed also in ref.~\cite{Birse:2005hh}.

\subsection{Simple pion emission from the nucleon}

First, in order to introduce some notations and to demonstrate our
approach on a very simple example, let us consider the emission of
a pion from the nucleon, i.e.~the amplitude $\M(N'\pi|N)$ defined
through\begin{equation}
\M(N(p')\,\pi^{a}(k)|N(p))\equiv\lim_{k^{2}\rightarrow m_{\pi}^{2}}\frac{k^{2}-m_{\pi}^{2}}{\i}\langle N(p')|\Phi^{a}|N(p)\rangle,\end{equation}
where $\Phi^{a}$ is the interpolating pion field\begin{equation}
\Phi^{a}\equiv\frac{\partial\cdot A^{a}}{f_{\pi}m_{\pi}^{2}},\qquad\langle0|\Phi^{a}|\pi^{b}\rangle=\delta^{ab},\label{PCAC}\end{equation}
$A_{\nu}^{a}$ the axial current, \begin{equation}
A_{\nu}^{a}=\bar{\psi}\gamma_{\nu}\gamma_{5}\frac{\tau^{a}}{2}\psi,\end{equation}
and (only within this subsection) $k$ denotes the nucleon momentum
difference,\begin{equation}
k^{\nu}=(p-p')^{\nu}.\end{equation}
If we relax somewhat about the on-shell requirement for the pion,
we can also write\begin{equation}
\langle N(p)\,\pi^{a}(k)|N(p)\rangle=(2\pi)^{4}\delta(p'+k-p)\,\M(N(p')\,\pi^{a}(k)|N(p)).\end{equation}

To derive the soft-pion theorem for the amplitude $\M(N'\pi|N)$,
we start from the nucleon matrix element\begin{equation}
I_{\nu}=\frac{1}{f_{\pi}}\langle N(p')|A_{\nu}^{a}|N(p)\rangle.\end{equation}
 Applying the definition of the interpolating pion field (\ref{PCAC})
yields the identity\begin{equation}
k^{\nu}I_{\nu}=\i m_{\pi}^{2}\langle N(p')|\Phi^{a}|N(p)\rangle.\label{k.I_1}\end{equation}
This equation is now investigated in the region where $k$ is small,\[
k\sim m_{\pi}\sim\varepsilon\]
i.e.~we search for the leading contributions in $\varepsilon$ to
both sides of equation (\ref{k.I_1}).

The nucleon matrix element on the left hand side is parametrized in
terms of the axial form factor $G_{A}$ and the pseudoscalar form
factor $G_{P}$,\begin{equation}
I_{\nu}=\frac{1}{f_{\pi}}\langle N'|A_{\nu}^{a}|N\rangle=\frac{1}{f_{\pi}}\bar{U}'\left[G_{A}(k^{2})\gamma_{\nu}-G_{P}(k^{2})\frac{k_{\nu}}{2M}\right]\gamma_{5}\frac{\tau^{a}}{2}U.\label{def_GA_GP}\end{equation}
For small $k^{2}$, $G_{A}$ approaches the axial coupling constant
and $G_{P}$ is dominated by the corresponding pion pole contribution,\begin{equation}
G_{A}(k^{2})=g_{A}+\mathcal{O}(\varepsilon^{2}),\qquad G_{P}(k^{2})=-\frac{(2M)^{2}g_{A}}{k^{2}-m_{\pi}^{2}}+\mathcal{O}(1/\varepsilon).\end{equation}
The insertion of these relations and the use of the Dirac equation
yield\begin{equation}
k^{\nu}I_{\nu}=-\frac{m_{\pi}^{2}}{k^{2}-m_{\pi}^{2}}\bar{U}'\frac{g_{A}}{2f_{\pi}}\slash{k}\gamma_{5}\tau^{a}U+\mathcal{O}(\varepsilon^{2}).\label{AxialCurrentInNucleon_approx}\end{equation}

The right hand side of equation (\ref{k.I_1}) has an explicit factor
of $m_{\pi}^{2}\sim\varepsilon^{2}$. This small factor can be compensated
only by the pion pole contribution,\begin{equation}
\i m_{\pi}^{2}\langle N(p')|\Phi^{a}|N(p)\rangle=\i m_{\pi}^{2}\frac{\i}{k^{2}-m_{\pi}^{2}}\M(N(p')\pi^{a}(k)|N(p))+\mathcal{O}(\varepsilon^{2}).\end{equation}
So finally, the equation (\ref{k.I_1}) leads to\begin{equation}
\M(N'\pi^{a}|N)=\bar{U}'\frac{g_{A}}{f_{\pi}}\slash k\gamma_{5}\frac{\tau^{a}}{2}U+\mathcal{O}(\varepsilon^{2}),\label{Goldberger_Treiman}\end{equation}
from which in turn the Goldberger-Treiman relation emerges, if we
parametrize $\M(N'\pi|N)$ in terms of the pion-nucleon coupling constant
$g_{\pi NN}$ as usual. The line of argument that lead to this well-known
result is now generalized to the case when a lightcone operator is
present.

\subsection{Pion emission induced by lightcone operators}

\subsubsection{Isovector quark operator of vector type}

Now we turn to the pion emission induced by lightcone operators. First,
we deal with the isovector quark operator of vector type which we
shall refer to as $O^{b}(\lambda)$ in the following:\begin{equation}
O^{b}(\lambda)\equiv\bar{\psi}(-\lambda n/2)\,\slash{n}\tau^{b}\psi(\lambda n/2).\end{equation}
We recall that in terms of local operators it reads\begin{equation}
O^{b}(\lambda)=\sum_{m=0}^{\infty}\frac{1}{m!}\left(\frac{\lambda}{2}\right)^{m}\bar{\psi}(0)\,[n\cdot(\overrightarrow{\partial}-\overleftarrow{\partial})]^{m}\slash n\tau^{b}\psi(0).\label{local_expansion}\end{equation}

The derivation of the soft-pion formula for $\langle N'\pi|O^{b}(\lambda)|N\rangle$
starts from the following object denoted $I_{\nu}$:\begin{equation}
I_{\nu}\equiv\frac{1}{f_{\pi}}\int\d^{4}z\,\e^{\i k\cdot z}\langle N(p')|T[A_{\nu}^{a}(z)\, O^{b}(\lambda)]|N(p)\rangle,\label{I_def}\end{equation}
where $T$ stands for the time-ordering prescription\begin{equation}
T[A_{\nu}^{a}(z)\, O^{b}(\lambda)]=\theta(z_{0})\, A_{\nu}^{a}(z)\, O^{b}(\lambda)+\theta(-z_{0})\, O^{b}(\lambda)\, A_{\nu}^{a}(z)\end{equation}
Let us investigate the behavior of the matrix element $I_{\nu}$ in
the soft-pion region\begin{equation}
k\sim m_{\pi}\sim\varepsilon.\label{soft_kinematics}\end{equation}
In this region, several momenta are close to the pion or nucleon mass
shell:\begin{equation}
k^{2}=m_{\pi}^{2}+\mathcal{O}(\varepsilon^{2}),\qquad W^{2}=(p'+k)^{2}=M^{2}+\mathcal{O}(\varepsilon),\qquad u=(p-k)^{2}=M^{2}+\mathcal{O}(\varepsilon).\label{mass_shell}\end{equation}
Generally, for a Green function of $n$ Operators,\begin{equation}
G_{1\ldots n}(\beta|\alpha)\equiv\int\d^{4}z_{1}\ldots\d^{4}z_{n-1}\e^{\i(p_{1}\cdot z_{1}+\ldots+p_{n-1}\cdot z_{n-1})}\langle\beta|T[O_{1}(z_{1})\ldots O_{n-1}(z_{n-1})\, O_{n}(0)]|\alpha\rangle,\end{equation}
we expect the following behavior when $|\alpha\rangle=|\beta\rangle=|0\rangle$
and the momentum $p\equiv p_{1}+\ldots+p_{r}$ reaches the mass shell
of a particle with mass $m$ and quantum numbers $s$:\begin{equation}
G_{1\ldots n}(0|0)\stackrel{(p_{1}+\ldots+p_{r})^{2}=p^{2}\rightarrow m^{2}}{\longrightarrow}\sum_{s}G_{1\ldots r}(0|p,s)\frac{\i}{p^{2}-m^{2}}G_{r+1\ldots n}(p,s|0)\label{general_pole}\end{equation}
(see, e.g., Chapter 10.2 in the textbook of Weinberg \cite{Weinberg96}).
The matrix element $I_{\nu}$ can be related to a Green function of
the type $G_{1\ldots4}(0|0)$ according to the usual LSZ formalism.
However, in our case we deal with the additional problem that \emph{several}
momentum combinations become nearly on-shell \emph{at} \emph{once},
\emph{}as shown in the equations (\ref{mass_shell}). We proceed by
adding up these different pole contributions, each of which is of
the type (\ref{general_pole}), and subtract appropriate terms where
necessary to avoid a double counting. In this way we obtain the preliminary
statement\begin{equation}
I^{\nu}=I_{\pi(k)}^{\nu}+I_{N(p'+k)}^{\nu}+I_{N(p-k)}^{\nu}+\mathcal{O}(\varepsilon^{0}),\qquad(\textrm{preliminary})\label{I_prelim}\end{equation}
where $I_{\pi(k)}^{\nu}$ denotes the pion pole in the variable $k^{2}$,\begin{equation}
I^{\nu}\stackrel{k^{2}\rightarrow m_{\pi}^{2}}{\longrightarrow}\frac{1}{f_{\pi}}\langle0|A^{\nu a}(0)|\pi^{c}(k)\rangle\frac{\i}{k^{2}-m_{\pi}^{2}}\langle N(p')\,\pi^{c}(k)|O^{b}(\lambda)|N(p)\rangle\equiv I_{\pi(k)}^{\nu}=\mathcal{O}(1/\varepsilon),\end{equation}
$I_{N(p'+k)}^{\nu}$ the nucleon pole in the variable $W^{2}$ with
pion-pole subtraction,\begin{eqnarray}
I^{\nu}-I_{\pi(k)}^{\nu} & \stackrel{W^{2}\rightarrow M^{2}}{\longrightarrow} & \frac{1}{f_{\pi}}\left[\langle N(p')|A^{\nu a}|N_{\sigma}(p'+k)\rangle\phantom{\frac{\i}{k^{2}-m_{\pi}^{2}}}\right.\nonumber \\
 &  & \left.-\langle0|A^{\nu a}|\pi^{c}(k)\rangle\frac{\i}{k^{2}-m_{\pi}^{2}}\M(N(p')\pi^{c}(k)|N_{\sigma}(p'+k))\right]\\
 &  & \times\frac{\i}{W^{2}-M^{2}}\langle N_{\sigma}(p'+k)|O^{b}(\lambda)|N(p)\rangle\equiv I_{N(p'+k)}^{\nu}=\mathcal{O}(1/\varepsilon),\end{eqnarray}
and $I_{N(p-k)}^{\nu}$ the nucleon pole in $u$ with pion-pole subtraction,\begin{eqnarray}
I^{\nu}-I_{\pi(k)}^{\nu} & \stackrel{u\rightarrow M^{2}}{\longrightarrow} & \frac{1}{f_{\pi}}\langle N(p')|O^{b}(\lambda)|N_{\sigma}(p-k)\rangle\frac{\i}{u-M^{2}}\left[\langle N_{\sigma}(p-k)|A^{\nu a}(0)|N(p)\rangle\phantom{\frac{\i}{k^{2}}}\right.\nonumber \\
 &  & \left.-\langle0|A^{\nu a}|\pi^{c}(k)\rangle\frac{\i}{k^{2}-m_{\pi}^{2}}\M(N_{\sigma}(p-k)\,\pi^{c}(k)|N(p))\right]\\
 &  & \equiv I_{N(p-k)}^{\nu}=\mathcal{O}(1/\varepsilon).\end{eqnarray}
In the previous formulas, $c$ denotes the pion's isospin and $\sigma$
a combined nucleon spin-isospin index. Summation over quantum numbers
of intermediate states, such as $c$ and $\sigma$ here, are henceforth
understood.

An additional pole that we have not mentioned yet has to be considered
in the kinematics of small momentum transfer, $t\sim\varepsilon^{2},$
since then\begin{equation}
t=m_{\pi}^{2}+\mathcal{O}(\varepsilon^{2}).\end{equation}
It has the form\begin{eqnarray}
I^{\nu}-I_{\pi(k)}^{\nu} & \stackrel{t\rightarrow m_{\pi}^{2}}{\longrightarrow} & \frac{1}{f_{\pi}}\left[\int\d^{4}z\,\e^{\i k\cdot z}\langle0|T[A^{\nu a}(z)\, O^{b}]|\pi^{d}(p-p')\rangle\right.\nonumber \\
 &  & \left.-\langle0|A^{\nu a}|\pi^{c}(k)\rangle\frac{\i}{k^{2}-m_{\pi}^{2}}\langle\pi^{c}(k)|O^{b}|\pi^{d}(p-p')\rangle\right]\\
 &  & \times\frac{\i}{t-m_{\pi}^{2}}\M(N'\pi^{d}(p-p')|N)\equiv I_{\pi(p-p')}^{\nu},\end{eqnarray}
where a $\pi(k)$-pole subtraction has been performed. Therefore,
to cover the whole region of momentum transfer $-t\lesssim M^{2}$,
we modify equation (\ref{I_prelim}) by adding this contribution,\begin{equation}
I^{\nu}=I_{\pi(k)}^{\nu}+I_{N(p'+k)}^{\nu}+I_{N(p-k)}^{\nu}+I_{\pi(p-p')}^{\nu}+\mathcal{O}(\varepsilon^{0}),\label{I_poles}\end{equation}
always keeping in mind that $I_{\pi(p-p')}^{\nu}$ is suppressed automatically
as $-t$ grows:\begin{equation}
I_{\pi(p-p')}^{\nu}=\left\{ \begin{array}{cc}
\mathcal{O}(1/\varepsilon) & t\sim\varepsilon^{2}\\
\mathcal{O}(\varepsilon^{0}) & -t\gg\varepsilon^{2}\end{array}\right..\end{equation}
This closes the discussion about the pole structure of the matrix
element $I^{\nu}$.

Next we turn to a simple identity for $I_{\nu}$ that follows when
we use the definition of the interpolating pion field (\ref{PCAC}):\begin{equation}
k^{\nu}I_{\nu}=\i\int\d^{4}z\,\e^{\i k\cdot z}\left\{ \frac{\delta(z_{0})}{f_{\pi}}\langle N'|[A_{0}^{a}(z),O^{b}(\lambda)]|N\rangle+m_{\pi}^{2}\langle N'|T[\Phi^{a}(z)\, O^{b}(\lambda)]|N\rangle\right\} .\label{k.I}\end{equation}
Let us exploit this identity in the soft-pion region (\ref{soft_kinematics}).
According to our previous demonstrations, the leading contributions
to the left hand side are simply obtained by contracting $k$ with
the poles in (\ref{I_poles}). The contraction with the $\pi(k)$
pole yields\begin{equation}
k_{\nu}I_{\pi(k)}^{\nu}=-\frac{k^{2}}{k^{2}-m_{\pi}^{2}}\langle N'\pi^{a}(k)|O^{b}(\lambda)|N\rangle+\mathcal{O}(\varepsilon),\end{equation}
where we have used the definition of the pion decay constant, \begin{equation}
\langle0|A_{\nu}^{a}|\pi^{b}(k)\rangle=\i f_{\pi}k_{\nu}\delta^{ab}.\end{equation}
For the sum of the nucleon poles we obtain\begin{eqnarray}
k_{\nu}(I_{N(p'+k)}^{\nu}+I_{N(p-k)}^{\nu}) & = & \bar{U}'\frac{g_{A}}{2f_{\pi}}\slash k\gamma_{5}\tau^{a}\frac{\i(\slash{p}'+M)}{2p'\cdot k}\Gamma^{(V)}(p'+k,p,\lambda)\,\tau^{b}U\nonumber \\
 &  & +\bar{U}'\Gamma^{(V)}(p',p-k,\lambda)\,\tau^{b}\frac{\i(\slash{p}+M)}{-2p\cdot k}\frac{g_{A}}{2f_{\pi}}\slash k\gamma_{5}\tau^{a}U+\mathcal{O}(\varepsilon),\end{eqnarray}
where we have applied\begin{equation}
\frac{k^{\nu}}{f_{\pi}}\langle N(p_{2})|A_{\nu}^{a}|N(p_{1}+k)\rangle=-\frac{m_{\pi}^{2}}{k^{2}-m_{\pi}^{2}}\bar{U}(p_{2})\frac{g_{A}}{2f_{\pi}}\slash k\gamma_{5}\tau^{a}U(p_{1})+\mathcal{O}(\varepsilon^{2})\end{equation}
and\begin{equation}
\M(N(p_{2})\,\pi^{a}(k)|N(p_{1}))=\bar{U}(p_{2})\frac{g_{A}}{2f_{\pi}}\slash k\gamma_{5}\tau^{a}U(p_{1})+\mathcal{O}(\varepsilon^{2}),\end{equation}
as discussed in the previous subsection. Moreover, we have introduced
the parametrization\begin{equation}
\langle N(p_{2})|O^{b}(\lambda)|N(p_{1})\rangle=\bar{U}(p_{2})\,\Gamma^{(V)}(p_{2},p_{1},\lambda)\,\tau^{b}\, U(p_{1}),\end{equation}
where in terms of generalized parton distributions one usually defines\begin{equation}
\Gamma^{(V)}(p_{2},p_{1},\lambda)\equiv\int\limits _{-1}^{1}\d x\,\e^{-\i\lambda x(p_{2}+p_{1})\cdot n/2}\left[H^{(V)}\slash n+E^{(V)}\frac{\i\sigma(n,p_{2}-p_{1})}{2M}\right],\label{GammaV}\end{equation}
with\begin{equation}
H^{(V)}=H^{(V)}\left(x,\frac{(p_{1}-p_{2})\cdot n}{(p_{1}+p_{2})\cdot n},(p_{1}-p_{2})^{2}\right)\label{GPD_arguments}\end{equation}
and analogously for $E^{(V)}$. Further, we mention that it is useful
to rewrite the contraction with the $\pi(p-p')$ pole in the form\begin{eqnarray}
\lefteqn{k_{\nu}I_{\pi(p-p')}^{\nu}}\nonumber \\
 & = & \left[\frac{\i}{f_{\pi}}\int\d^{4}z\,\e^{\i k\cdot z}\{\delta(z_{0})\langle0|[A_{0}^{a}(z),O^{b}]|\pi^{d}(p-p')\rangle+m_{\pi}^{2}\langle0|T[\Phi^{a}(z)\, O^{b}]|\pi^{d}(p-p')\rangle\}\right.\nonumber \\
 &  & \left.-\frac{k_{\nu}}{f_{\pi}}\langle0|A^{\nu a}|\pi^{c}(k)\rangle\frac{\i}{k^{2}-m_{\pi}^{2}}\langle\pi^{c}(k)|O^{b}|\pi^{d}(p-p')\rangle\right]\frac{\i}{t-m_{\pi}^{2}}\M(N'\,\pi^{d}(p-p')|N)\\
 & = & \left[\frac{\i}{f_{\pi}}\langle0|[Q_{5}^{a}(0),O^{b}]|\pi^{d}(p-p')\rangle+\i m_{\pi}^{2}\int\d^{4}z\,\e^{\i k\cdot z}\langle0|T[\Phi^{a}(z)\, O^{b}]|\pi^{d}(p-p')\rangle\right.\nonumber \\
 &  & \left.+\frac{k^{2}}{k^{2}-m_{\pi}^{2}}\langle\pi^{a}(k)|O^{b}|\pi^{d}(p-p')\rangle\right]\frac{\i}{t-m_{\pi}^{2}}\bar{U}'\frac{g_{A}}{2f_{\pi}}(\slash p-\slash p')\gamma_{5}\tau^{d}U+\mathcal{O}(\varepsilon),\end{eqnarray}
where $Q_{5}^{a}$ is the axial charge,\begin{equation}
Q_{5}^{a}(z_{0})\equiv\int\d^{3}z\, A_{0}^{a}(z).\end{equation}
So much for the left hand side of the identity (\ref{k.I}). Now,
what are the main terms on the right hand side?

First, we turn to the commutator term. It can be approximated by neglecting
the soft momentum $k$,\begin{equation}
\i\int\d^{4}z\,\e^{\i k\cdot z}\frac{\delta(z_{0})}{f_{\pi}}\langle N'|[A_{0}^{a}(z),O^{b}(\lambda)]|N\rangle=\frac{\i}{f_{\pi}}\langle N'|[Q_{5}^{a}(0),O^{b}(\lambda)]|N\rangle+\mathcal{O}(\varepsilon).\end{equation}
For the calculation of the commutator $[Q_{5}^{a}(0),O^{b}(\lambda)]$,
we insert $O^{b}(\lambda)$ in the form (\ref{local_expansion}).
Then we have to deal with objects of the type $[Q_{5}^{a}(0),(n\cdot\partial)^{m}\psi(0)]$
and $[Q_{5}^{a}(0),(n\cdot\partial)^{m}\bar{\psi}(0)]$. For $m=0$,
these commutators are well-known from the transformation of the fields
under the axial part of the chiral rotation:\begin{equation}
[Q_{5}^{a}(0),\,\psi(0)]=-\gamma_{5}\frac{\tau^{a}}{2}\psi(0),\qquad[Q_{5}^{a}(0),\bar{\psi}(0)]=-\bar{\psi}(0)\gamma_{5}\frac{\tau^{a}}{2}.\end{equation}
In the general case $m\ge0$, we can rewrite e.g.\begin{equation}
[Q_{5}^{a},(n\cdot\partial)^{m}\psi]=\sum_{k=0}^{m}\left(\begin{array}{c}
m\\
k\end{array}\right)(-1)^{k}(n\cdot\partial)^{m-k}[(n_{0}\partial_{0})^{k}Q_{5}^{a},\psi].\end{equation}
The derivative of the axial charge is given as\begin{equation}
\partial_{0}Q_{5}^{a}(z_{0})=\int\d^{3}z\,\bar{\psi}(z)\,\gamma_{5}\{\tau^{a},\hat{m}\}\psi(z),\end{equation}
where $\hat{m}=\textrm{diag}(m_{u},m_{d})$ is the quark mass matrix.
According to the Gell-Mann-Oakes-Renner relation we have \[
m_{u},m_{d}\propto m_{\pi}^{2}\sim\varepsilon^{2},\]
hence we can neglect the terms that involve derivatives of $Q_{5}^{a}$:\begin{equation}
[Q_{5}^{a},(n\cdot\partial)^{m}\psi]=(n\cdot\partial)^{m}[Q_{5}^{a},\psi]+\mathcal{O}(\varepsilon^{2})\end{equation}
and similar for $[Q_{5}^{a},(n\cdot\partial)^{m}\bar{\psi}]$. In
this way we obtain finally\begin{equation}
[Q_{5}^{a}(0),O^{b}(\lambda)]=\i\varepsilon^{abc}O_{5}^{c}(\lambda)+\mathcal{O}(\varepsilon^{2})\end{equation}
with\begin{equation}
O_{5}^{c}(\lambda)\equiv\bar{\psi}(-\lambda n/2)\,\slash n\gamma_{5}\tau^{c}\psi(\lambda n/2).\end{equation}

Next, we discuss the second term on the right hand side of (\ref{k.I}).
It is accompanied by a small factor $m_{\pi}^{2}\sim\varepsilon^{2}$,
so again we focus on the pole contributions which are of order $1/\varepsilon^{2}$.
These are the pion pole in $k^{2}$ and, for small momentum transfer,
the pion pole in $t$. Again taking into account a double-pole subtraction,
we thus get\begin{eqnarray}
\lefteqn{\i m_{\pi}^{2}\int\d^{4}z\,\e^{\i k\cdot z}\langle N'|T[\Phi^{a}(z)\, O^{b}]|N\rangle}\nonumber \\
 & = & -\frac{m_{\pi}^{2}}{k^{2}-m_{\pi}}\langle N'\pi^{a}(k)|O^{b}|N\rangle+\left\{ \i m_{\pi}^{2}\int\d^{4}z\,\e^{\i k\cdot z}\langle0|T[\Phi^{a}(z)\, O^{b}]|\pi^{d}(p-p')\rangle\right.\nonumber \\
 &  & \left.+\frac{m_{\pi}^{2}}{k^{2}-m_{\pi}^{2}}\langle\pi^{a}(k)|O^{b}(\lambda)|\pi^{d}(p-p')\rangle\right\} \frac{\i}{t-m_{\pi}^{2}}\M(N'\pi^{d}(p-p')|N)+\mathcal{O}(\varepsilon).\end{eqnarray}

\begin{figure}[!t]
\begin{center}\includegraphics[%
  width=0.90\textwidth]{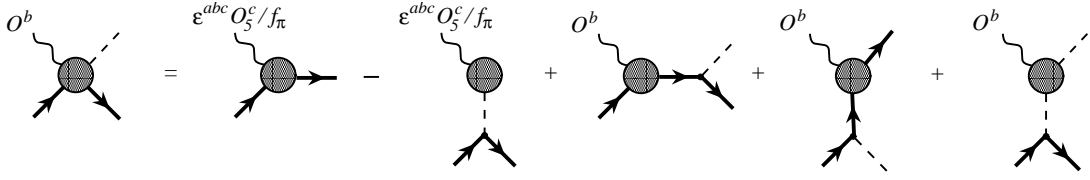}\end{center}

\caption{\label{figure_SPT} Illustration of the soft-pion theorem for the
matrix element $\langle N(p')\,\pi^{a}(k)|O^{b}(\lambda)|N(p)\rangle$,
where $O^{b}(\lambda)$ is the lightcone operator $\bar{\psi}(-\lambda n/2)\,\slash n\tau^{b}\psi(\lambda n/2).$
The blobs denote insertion of the indicated operator while the pointlike
vertices represent the standard pseudovector pion-nucleon coupling.}\lyxline{\normalsize}

\end{figure}

Now we can collect all approximations and insert them into (\ref{k.I})
(note that the terms which involve the time-orderd product $T[\Phi^{a}(z),O^{b}]$
cancel). Then we can solve the equation for the matrix element $\langle N'\pi^{a}|O^{b}|N\rangle$
and arrive at the soft-pion theorem\begin{eqnarray}
\lefteqn{\langle N'\pi^{a}(k)|O^{b}(\lambda)|N\rangle}\nonumber \\
 & = & \frac{\varepsilon^{abc}}{f_{\pi}}\left[\langle N'|O_{5}^{c}(\lambda)|N\rangle-\langle0|O_{5}^{c}(\lambda)|\pi^{d}(p-p')\rangle\frac{\i}{t-m_{\pi}^{2}}\bar{U}'\frac{g_{A}}{2f_{\pi}}(\slash{p}-\slash{p}')\gamma_{5}\tau^{d}U\right]\nonumber \\
 &  & +\frac{\i g_{A}}{2f_{\pi}}\bar{U}'\left[\slash k\gamma_{5}\tau^{a}\frac{\slash{p}'+M}{2p'\cdot k}\Gamma^{(V)}(p'+k,p,\lambda)\,\tau^{b}+\Gamma^{(V)}(p',p-k,\lambda)\,\tau^{b}\frac{\slash{p}+M}{-2p\cdot k}\slash k\gamma_{5}\tau^{a}\right]U\nonumber \\
 &  & +\langle\pi^{a}(k)|O^{b}(\lambda)|\pi^{d}(p-p')\rangle\frac{\i}{t-m_{\pi}^{2}}\bar{U}'\frac{g_{A}}{2f_{\pi}}(\slash{p}-\slash{p}')\gamma_{5}\tau^{d}U+\mathcal{O}(\varepsilon),\label{SPT}\end{eqnarray}
see also figure \ref{figure_SPT}.%
 The pion matrix elements in this formula are parametrized as\[
\langle0|O_{5}^{c}(\lambda)|\pi^{d}(p-p')\rangle=2\i f_{\pi}(p-p')\cdot n\,\delta^{cd}\int\limits _{-1}^{1}\d x\,\e^{-\i\lambda x(p-p')\cdot n/2}\phi_{\pi}(x),\]
where $\phi_{\pi}$ is the pion distribution amplitude, and\begin{eqnarray*}
\lefteqn{\langle\pi^{a}(k)|O^{b}(\lambda)|\pi^{d}(p-p')\rangle}\\
 &  & =\i\varepsilon^{abd}(k+p-p')\cdot n\int\limits _{-1}^{1}\d x\,\e^{-\i\lambda x(k+p-p')\cdot n/2}H_{\pi}^{(V)}\left(x,\frac{(p-p'-k)\cdot n}{(p-p'+k)\cdot n},0\right)+\mathcal{O}(\varepsilon^{3}),\quad\end{eqnarray*}
In the last line, we have neglected the momentum transfer in the argument
of the pion GPD $H_{\pi}^{(V)}$, because the $\pi(p-p')$ pole terms
only contribute in the region where $t\sim\Delta^{2}\sim\varepsilon^{2}$.
For a parametrization of the nucleon matrix element of $O_{5}^{b}(\lambda)$
in terms of GPDs see below, formulas (\ref{Gamma_5_def0}) and (\ref{Gamma_5_def}).

For reasons of consistency, it is advisable to check that the $\pi(p-p')$
pole contributions in the soft-pion formula is really negligible for
$-t\gg\varepsilon^{2}$, as we have promised at the beginning of the
calculation. In this region of moderate momentum transfer, we have
to distinguish two cases which are both kinematically allowed:\begin{equation}
(p-p')\cdot n\sim k\cdot n\sim\varepsilon\qquad\textrm{or}\qquad(p-p')\cdot n\gg k\cdot n\sim\varepsilon.\end{equation}
In the first case, both pion pole terms are individually suppressed
because the pion matrix elements are proportional to combinations
of $(p-p')\cdot n$ and $k\cdot n$. In the latter case, the \emph{combination}
of both terms is small because\begin{eqnarray}
 &  & (k+p-p')\cdot n\int\limits _{-1}^{1}\d x\,\e^{-\i\lambda x(k+p-p')\cdot n/2}H_{\pi}^{(V)}\left(x,\frac{(p-p'-k)\cdot n}{(p-p'+k)\cdot n},0\right)\nonumber \\
 & \stackrel{(p-p')\cdot n\gg k\cdot n\sim\varepsilon}{=} & (p-p')\cdot n\int\limits _{-1}^{1}\d x\,\e^{-\i\lambda x(p-p')\cdot n/2}H_{\pi}^{(V)}(x,1,0)+\mathcal{O}(\varepsilon)\end{eqnarray}
and a soft-pion theorem for the pion GPD \cite{Polyakov:1998ze} reads\begin{equation}
H_{\pi}^{(V)}(x,1,0)=2\phi_{\pi}(x).\end{equation}
So we have confirmed that in the region of moderate momentum transfer
the $\pi(p-p')$ pole is negligible. Therefore, our result coincides
with that one of Guichon et.~al.~\cite{Guichon:2003ah}, who restricted
themselves to the region $-t\gg\varepsilon^{2}$.

\subsubsection{Isocalar quark operator of vector type and gluon operator}

Soft-pion theorems similar to the previous one can also be derived
for the isoscalar quark operator and the gluon operator. For the isoscalar
quark operator, we get\begin{eqnarray}
\lefteqn{\langle N'\pi^{a}(k)|\bar{\psi}(-\lambda n/2)\slash{n}\psi(\lambda n/2)|N\rangle}\nonumber \\
 & = & \frac{\i g_{A}}{2f_{\pi}}\bar{U}'\left[\slash k\gamma_{5}\tau^{a}\frac{\slash p'+M}{2p'\cdot k}\Gamma^{(S)}(p'+k,p,\lambda)-\Gamma^{(S)}(p',p-k,\lambda)\frac{\slash p+M}{2p\cdot k}\slash k\gamma_{5}\tau^{a}\right]U\nonumber \\
 &  & +\langle\pi^{a}(k)|\bar{\psi}(-\lambda n/2)\slash n\psi(\lambda n/2)|\pi^{c}(p-p')\rangle\frac{\i}{t-m_{\pi}^{2}}\bar{U}'\frac{g_{A}}{2f_{\pi}}(\slash p-\slash p')\gamma_{5}\tau^{c}U+\mathcal{O}(\varepsilon),\quad\label{SPT_vector_isoscalar}\end{eqnarray}
where $\Gamma^{(S)}$ can be obtained from the definition of $\Gamma^{(V)}$
in equation (\ref{GammaV}) by replacing the isovector GPDs $H^{(V)}$
and $E^{(V)}$ with the isoscalar ones, $H^{(S)}$ and $E^{(S)}$.
For the gluon operator, we obtain\begin{eqnarray}
\lefteqn{\langle N'\pi^{a}(k)|n_{\mu}F^{\mu\rho}(-\lambda n/2)\, F_{\rho\nu}(\lambda n/2)\, n^{\nu}|N\rangle}\nonumber \\
 & = & \frac{\i g_{A}}{2f_{\pi}}\bar{U}'\left[\slash k\gamma_{5}\tau^{a}\frac{\slash p'+M}{2p'\cdot k}\Gamma^{(G)}(p'+k,p,\lambda)-\Gamma^{(G)}(p',p-k,\lambda)\frac{\slash p+M}{2p\cdot k}\slash k\gamma_{5}\tau^{a}\right]U\nonumber \\
 &  & +\langle\pi^{a}(k)|n_{\mu}F^{\mu\rho}(-\lambda n/2)\, F_{\rho\nu}(\lambda n/2)\, n^{\nu}|\pi^{c}(p-p')\rangle\frac{\i}{t-m_{\pi}^{2}}\bar{U}'\frac{g_{A}}{2f_{\pi}}(\slash p-\slash p')\gamma_{5}\tau^{c}U\nonumber \\
 &  & +\mathcal{O}(\varepsilon),\label{SPT_gluon}\end{eqnarray}
where\begin{equation}
\Gamma^{(G)}(p_{2},p_{1},\lambda)=\frac{n\cdot(p_{1}+p_{2})}{4}\int\limits _{-1}^{1}\d x\,\e^{-\i\lambda x\, n\cdot(p_{1}+p_{2})/2}x\left[H^{(G)}\slash{n}+E^{(G)}\frac{\i\sigma(n,p_{2}-p_{1})}{2M}\right],\end{equation}
with arguments of the gluon GPDs $H^{(G)}$ and $E^{(G)}$ as in (\ref{GPD_arguments}).
Because of the isoscalar nature of the two operators considered here,
commutator terms do not appear in the results. Moreover, we remark
that the discussion of the pion pole terms at moderate momentum transfer
which we presented in the isovector case, can be repeated here in
a similar way using the following soft-pion theorems for the gluon
and the isoscalar pion GPD \cite{Polyakov:1998ze}:\begin{equation}
H^{(S)}(x,1,0)=0,\qquad H^{(G)}(x,1,0)=0.\end{equation}
So again, for $-t\gg\varepsilon^{2}$ the pion pole contributions
to the soft-pion theorem are negligible.

\subsubsection{Isovector operator of axial vector type}

Next we turn to the derivation of the soft-pion theorem for the matrix
element of the operator\begin{equation}
O_{5}^{b}(\lambda)=\bar{\psi}(-\lambda n/2)\slash{n}\gamma_{5}\tau^{b}\psi(\lambda n/2).\end{equation}
For this purpose, we define\begin{equation}
I^{\nu}\equiv\frac{1}{f_{\pi}}\int\d^{4}z\,\e^{\i k\cdot z}\langle N(p')|T[A_{\nu}^{a}(z)\, O_{5}^{b}(\lambda)]|N(p)\rangle\end{equation}
and obtain the identity \begin{equation}
k_{\nu}I^{\nu}=\i\int\d^{4}z\,\e^{\i k\cdot z}\left\{ \frac{\delta(z_{0})}{f_{\pi}}\langle N'|[A_{0}^{a}(z),O_{5}^{b}(\lambda)]|N\rangle+m_{\pi}^{2}\langle N'|T[\Phi^{a}(z)\, O_{5}^{b}(\lambda)]|N\rangle\right\} .\label{k.I5}\end{equation}
Again, we make soft-pion expansions of both sides when $k\sim m_{\pi}\sim\varepsilon$.
At this, we have to consider pion poles in $k^{2}$ and $\Delta^{2}=(p'+k-p)^{2}$
and nucleon poles:\begin{equation}
I^{\nu}=I_{\pi(k)}^{\nu}+I_{\pi(\Delta)}^{\nu}+I_{N(p'+k)}^{\nu}+I_{N(p-k)}^{\nu}+\mathcal{O}(\varepsilon^{0}).\end{equation}
The $\pi(k)$ pole is given as\begin{eqnarray}
I^{\nu}\stackrel{k^{2}\rightarrow m_{\pi}^{2}}{\longrightarrow} & = & \frac{1}{f_{\pi}}\langle0|A^{\nu a}(0)|\pi^{c}(k)\rangle\frac{\i}{k^{2}-m_{\pi}^{2}}\langle N'\pi^{c}(k)|O_{5}^{b}(\lambda)|N\rangle\\
 & = & \frac{-k^{\nu}}{k^{2}-m_{\pi}^{2}}\langle N'\pi^{a}(k)|O_{5}^{b}(\lambda)|N\rangle\equiv I_{\pi(k)}^{\nu}\end{eqnarray}
For the $\pi(\Delta)$ pole we get \begin{eqnarray}
I^{\nu}-I_{\pi(k)}^{\nu} & \stackrel{\Delta^{2}\rightarrow m_{\pi}^{2}}{\longrightarrow} & =\left[\frac{1}{f_{\pi}}\langle N'|A^{\nu a}(0)|N\pi^{c}(\Delta)\rangle+\frac{k^{\nu}}{k^{2}-m_{\pi}^{2}}\M(N'\pi^{a}(k)|N\pi^{c}(\Delta))\right]\nonumber \\
 &  & \times\frac{\i}{\Delta^{2}-m_{\pi}^{2}}\langle\pi^{c}(\Delta)|O^{b}(\lambda)|0\rangle\equiv I_{\pi(\Delta)}^{\nu},\end{eqnarray}
where $\M(N'\pi^{d}(k)|N\pi^{c}(\Delta))$ denotes the pion-nucleon
scattering amplitude,\begin{equation}
\langle N'\pi^{d}(k)|N\pi^{c}(\Delta)\rangle=(2\pi)^{4}\delta(p'+k-p-\Delta)\,\M(N'\pi^{d}(k)|N\pi^{c}(\Delta)).\end{equation}
This pion pole is non-negliglible only if $-\Delta^{2}\lesssim m_{\pi}^{2}\sim\varepsilon^{2}$.
The nucleon pole terms are\begin{eqnarray}
\lefteqn{I^{\nu}-I_{\pi(k)}^{\nu}-I_{\pi(\Delta)}^{\nu}}\\
 & \stackrel{W^{2}\rightarrow M^{2}}{\longrightarrow} & \left[\frac{1}{f_{\pi}}\langle N'|A^{\nu a}|N_{\sigma}(p'+k)\rangle+\frac{k^{\nu}}{k^{2}-m_{\pi}^{2}}\M(N'\pi^{a}(k)|N_{\sigma}(p'+k))\right]\frac{\i}{W^{2}-M^{2}}\nonumber \\
 &  & \times\left[\langle N_{\sigma}(p'+k)|O_{5}^{b}|N\rangle-\M(N_{\sigma}(p'+k)|N\pi^{c}(\Delta))\frac{\i}{\Delta^{2}-m_{\pi}^{2}}\langle\pi^{c}(\Delta)|O_{5}^{b}|0\rangle\right]\qquad\\
 & = & \bar{U}'\frac{\i g_{A}}{2f_{\pi}}\gamma^{\nu}\gamma_{5}\tau^{a}\frac{\slash p'+M}{2p'\cdot k}\left[\Gamma_{5}^{(V)}(p'+k,p,\lambda)\,\tau^{b}+\frac{\i g_{A}}{2f_{\pi}}\slash\Delta\gamma_{5}\tau^{c}\frac{\langle\pi^{c}(\Delta)|O_{5}^{b}|0\rangle}{\Delta^{2}-m_{\pi}^{2}}\right]U\nonumber \\
 &  & +\mathcal{O}(\varepsilon^{0})\equiv I_{N(p'+k)}^{\nu}\label{I_N(pp+k)}\end{eqnarray}
and\begin{eqnarray}
\lefteqn{I^{\nu}-I_{\pi(k)}^{\nu}-I_{\pi(\Delta)}^{\nu}}\\
 & \stackrel{u\rightarrow M^{2}}{\longrightarrow} & \left[\langle N'|O_{5}^{b}|N_{\sigma}(p-k)\rangle-\M(N'|N_{\sigma}(p-k)\pi^{c}(\Delta))\frac{\i}{\Delta^{2}-m_{\pi}^{2}}\langle\pi^{c}(\Delta)|O_{5}^{b}|0\rangle\right]\nonumber \\
 &  & \times\frac{\i}{u-M^{2}}\left[\frac{1}{f_{\pi}}\langle N_{\sigma}(p-k)|A^{\nu a}|N\rangle+\frac{k^{\nu}}{k^{2}-m_{\pi}^{2}}\M(N_{\sigma}(p-k)\pi^{a}(k)|N)\right]\\
 & = & \bar{U}'\left[\Gamma_{5}^{(V)}(p',p-k)\,\tau^{b}+\frac{\i g_{A}}{2f_{\pi}}\slash\Delta\gamma_{5}\tau^{c}\frac{\langle\pi^{c}(\Delta)|O_{5}^{b}|0\rangle}{\Delta^{2}-m_{\pi}^{2}}\right]\frac{\slash p+M}{-2p\cdot k}\frac{\i g_{A}}{2f_{\pi}}\gamma^{\nu}\gamma_{5}\tau^{a}U\nonumber \\
 &  & +\mathcal{O}(\varepsilon^{0})\equiv I_{N(p-k)}^{\nu},\label{I_N(p-k)}\end{eqnarray}
where we make use of the parametrization\begin{equation}
\langle N(p_{2})|O_{5}^{b}(\lambda)|N(p_{1})\rangle=\bar{U}(p_{2})\,\Gamma_{5}^{(V)}(p_{2},p_{1},\lambda)\,\tau^{b}U(p_{1})\label{Gamma_5_def0}\end{equation}
with\begin{equation}
\Gamma_{5}^{(V)}(p_{2},p_{1},\lambda)=\int\limits _{-1}^{1}\d x\,\e^{-\i\lambda x(p_{1}+p_{2})\cdot n/2}\left[\tilde{H}^{(V)}\slash n+\tilde{E}^{(V)}\frac{(p_{2}-p_{1})\cdot n}{(2M)^{2}}(\slash p_{2}-\slash p_{1})\right]\gamma_{5}.\label{Gamma_5_def}\end{equation}
Note that in contrast to the previous case of the vector type operator
$O^{b}(\lambda)$, a $\pi(\Delta)$ pole subtraction has been necessary
to obtain the correct nucleon pole terms (\ref{I_N(pp+k)}) and (\ref{I_N(p-k)})
without double counting.

From the soft-pion expansion for $I^{\nu}$ that we have determined
now, we immediately obtain the left hand side of the identity (\ref{k.I5}),
$k_{\nu}I^{\nu}$. The right hand side gives\begin{eqnarray}
 &  & \i\int\d^{4}z\,\e^{\i k\cdot z}\left\{ \frac{\delta(z_{0})}{f_{\pi}}\langle N'|[A_{0}^{a}(z),O_{5}^{b}(\lambda)]|N\rangle+m_{\pi}^{2}\langle N'|T[\Phi^{a}(z)\, O_{5}^{b}(\lambda)]|N\rangle\right\} \nonumber \\
 & = & -\frac{\varepsilon^{abc}}{f_{\pi}}\langle N'|O^{c}(\lambda)|N\rangle-\frac{m_{\pi}^{2}}{k^{2}-m_{\pi}^{2}}\langle N'\pi^{a}(k)|O_{5}^{b}(\lambda)|N\rangle\nonumber \\
 &  & +\left[\frac{k^{\nu}}{f_{\pi}}\langle N'|A_{\nu}^{a}|N\pi^{c}(\Delta)\rangle+\frac{m_{\pi}^{2}}{k^{2}-m_{\pi}^{2}}\M(N'\pi^{a}(k)|N\pi^{c}(\Delta))\right]\frac{\i}{\Delta^{2}-m_{\pi}^{2}}\langle\pi^{c}(\Delta)|O_{5}^{b}(\lambda)|0\rangle\nonumber \\
 &  & +\mathcal{O}(\varepsilon).\end{eqnarray}
In this way, we obtain from (\ref{k.I5}) the following first version
of the soft-pion theorem: \begin{eqnarray}
\lefteqn{\langle N'\pi^{a}(k)|O_{5}^{b}(\lambda)|N\rangle}\nonumber \\
 & = & \frac{\varepsilon^{abc}}{f_{\pi}}\langle N'|O^{c}(\lambda)|N\rangle+\M(N'\pi^{a}(k)|N\pi^{c}(\Delta))\frac{\i}{\Delta^{2}-m_{\pi}^{2}}\langle\pi^{c}(\Delta)|O_{5}^{b}(\lambda)|0\rangle\nonumber \\
 &  & +\bar{U}'\frac{\i g_{A}}{2f_{\pi}}\slash k\gamma_{5}\tau^{a}\frac{\slash{p'}+M}{2p'\cdot k}\left[\Gamma_{5}^{(V)}(p'+k,p,\lambda)\,\tau^{b}+\frac{\i g_{A}}{2f_{\pi}}\slash{\Delta}\gamma_{5}\tau^{c}\frac{\langle\pi^{c}(\Delta)|O_{5}^{b}(\lambda)|0\rangle}{\Delta^{2}-m_{\pi}^{2}}\right]U\nonumber \\
 &  & -\bar{U}'\left[\Gamma_{5}^{(V)}(p',p-k,\lambda)\,\tau^{b}+\frac{\i g_{A}}{2f_{\pi}}\slash{\Delta}\gamma_{5}\tau^{c}\frac{\langle\pi^{c}(\Delta)|O_{5}^{b}(\lambda)|0\rangle}{\Delta^{2}-m_{\pi}^{2}}\right]\frac{\slash{p}+M}{2p\cdot k}\frac{\i g_{A}}{2f_{\pi}}\slash k\gamma_{5}\tau^{a}U\nonumber \\
 &  & +\mathcal{O}(\varepsilon).\label{SPT5_preliminary}\end{eqnarray}

\begin{figure}[!t]
\begin{center}\includegraphics[%
  width=0.90\textwidth]{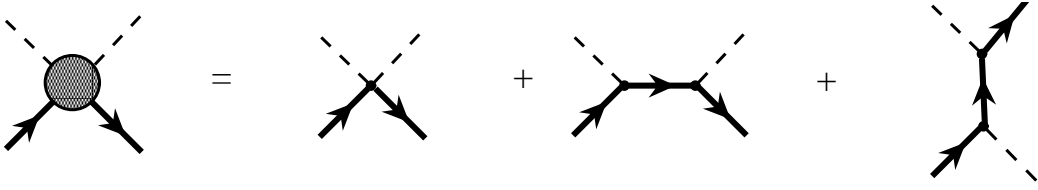}\end{center}

\caption{\label{figure_piN_scattering} Leading contributions to the pion-nucleon
scattering amplitude $\M(N(p')\,\pi(\Delta)|N(p)\,\pi(k))$ for soft
pion momenta $\Delta$ and $k$. The pointlike vertex with four particles
attached corresponds to the Weinberg coupling.}\lyxline{\normalsize}

\end{figure}
\begin{figure}[!t]
\begin{center}\includegraphics[%
  width=0.95\textwidth]{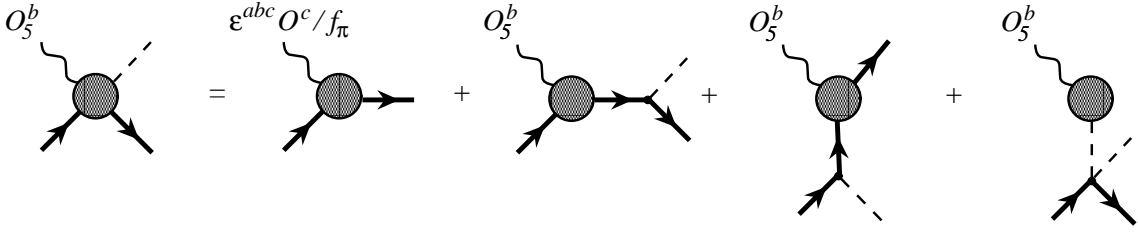}\end{center}

\caption{\label{figure_SPT5} The soft-pion theorem for the matrix element
$\langle N(p')\,\pi^{a}(k)|O_{5}^{b}(\lambda)|N(p)\rangle$, where
$O_{5}^{b}(\lambda)$ is the operator $\bar{\psi}(-\lambda n/2)\,\slash n\gamma_{5}\tau^{b}\psi(\lambda n/2)$. }\lyxline{\normalsize}

\end{figure}

To further simplify this expression, we need the pion-nucleon scattering
amplitude which appears in the $\pi(\Delta)$ pole terms in the region
of small momentum transfer. We can derive the soft-pion theorem for
this amplitude from the identity\begin{equation}
\frac{\Delta^{\mu}}{f_{\pi}}\langle N'\pi^{a}(k)|A_{\mu}^{c}(0)|N\rangle=-\i m_{\pi}^{2}\langle N'\pi^{a}(k)|\Phi^{c}(0)|N\rangle.\label{piN_scattering_derivation}\end{equation}
For the matrix element $\langle N'\pi^{a}|A_{\mu}^{c}|N\rangle$ on
the left hand side, we can use the soft-pion theorem (\ref{SPT5_preliminary})
since \begin{equation}
A_{\mu}^{c}(0)=2O_{5}^{c}(\lambda=0).\end{equation}
The right hand side of (\ref{piN_scattering_derivation}) is dominated
by the pion pole in $\Delta^{2}$,\begin{equation}
-\i m_{\pi}^{2}\langle N'\pi^{a}(k)|\Phi^{c}(0)|N\rangle=\M(N'\pi^{a}(k)|N\pi^{c}(\Delta))\frac{m_{\pi}^{2}}{\Delta^{2}-m_{\pi}^{2}}+\mathcal{O}(\varepsilon^{2}).\end{equation}
Thus, we can solve equation (\ref{piN_scattering_derivation}) for
the pion scattering amplitude and obtain the well-known result\begin{eqnarray}
\lefteqn{\M(N'\pi^{a}(k)|N\pi^{c}(\Delta))=-\frac{\varepsilon^{acd}}{4f_{\pi}^{2}}\bar{U}'(\slash k+\slash\Delta)\tau^{d}U}\nonumber \\
 &  & -\left(\frac{g_{A}}{f_{\pi}}\right)^{2}\bar{U}'\left(\slash k\gamma_{5}\frac{\tau^{a}}{2}\i\frac{\slash p'+M}{2p'\cdot k}\slash\Delta\gamma_{5}\frac{\tau^{c}}{2}-\slash\Delta\gamma_{5}\frac{\tau^{c}}{2}\i\frac{\slash p+M}{2p\cdot k}\slash k\gamma_{5}\frac{\tau^{a}}{2}\right)U+\mathcal{O}(\varepsilon^{2}),\end{eqnarray}
see also Fig. \ref{figure_piN_scattering}. If we insert this expression
into equation (\ref{SPT5_preliminary}), we find that the double poles
are canceled, and the final version of the soft-pion theorem reads\begin{eqnarray}
\lefteqn{\langle N'\pi^{a}(k)|O_{5}^{b}(\lambda)|N\rangle}\nonumber \\
 & = & \frac{\varepsilon^{abc}}{f_{\pi}}\bar{U}'\Gamma^{(V)}(p',p,\lambda)\,\tau^{c}U-\frac{\varepsilon^{acd}}{4f_{\pi}^{2}}\bar{U}'(\slash k+\slash\Delta)\tau^{d}U\frac{\i}{\Delta^{2}-m_{\pi}^{2}}\langle\pi^{c}(\Delta)|O_{5}^{b}(\lambda)|0\rangle\nonumber \\
 &  & +\frac{\i g_{A}}{2f_{\pi}}\bar{U}'\left[\slash k\gamma_{5}\tau^{a}\frac{\slash{p'}+M}{2p'\cdot k}\Gamma_{5}^{(V)}(p'+k,p,\lambda)\,\tau^{b}-\Gamma_{5}^{(V)}(p',p-k,\lambda)\,\tau^{b}\frac{\slash{p}+M}{2p\cdot k}\slash k\gamma_{5}\tau^{a}\right]U\nonumber \\
 &  & +\mathcal{O}(\varepsilon),\label{SPT5_isovector}\end{eqnarray}
see also figure \ref{figure_SPT5}.%

\paragraph{Moderate momentum transfer}

In the region of moderate momentum transfer, i.e.~$-\Delta^{2}$
$\gg\varepsilon^{2}$, the pion pole term in the soft-pion theorem
(\ref{SPT5_isovector}) is negligible as it should since according
to the Dirac equation for the nucleon spinors we have the suppression
factor\begin{equation}
\bar{U}'(\slash k+\slash\Delta)U=2\bar{U}'\slash kU=\mathcal{O}(\varepsilon),\end{equation}
while the denominator $\Delta^{2}-m_{\pi}^{2}$ is no longer small.
Therefore, in this region the soft-pion theorem again agrees with
the corresponding one in Guichon et~al.~\cite{Guichon:2003ah}.

\subsubsection{Isoscalar operator of axial vector type}

In the case of the isoscalar axial quark operator, matters are simpler
because commutator and pion pole terms vanish and we are simply left
with the nucleon pole terms:\begin{eqnarray}
\lefteqn{\langle N'\pi^{a}(k)|\bar{\psi}(-\lambda n/2)\slash n\gamma_{5}\psi(\lambda n/2)|N\rangle}\nonumber \\
 & = & \frac{\i g_{A}}{2f_{\pi}}\bar{U}'\left[\slash k\gamma_{5}\tau^{a}\frac{\slash p'+M}{2p'\cdot k}\Gamma_{5}^{(S)}(p'+k,p,\lambda)-\Gamma_{5}^{(S)}(p',p-k,\lambda)\frac{\slash p+M}{2p\cdot k}\slash k\gamma_{5}\tau^{a}\right]U,\nonumber \\
 &  & +\mathcal{O}(\varepsilon),\label{SPT5_isoscalar}\end{eqnarray}
where\begin{equation}
\Gamma_{5}^{(S)}(p_{2},p_{1},\lambda)=\int\limits _{-1}^{1}\d x\,\e^{-\i\lambda x\, n\cdot(p_{1}+p_{2})/2}\left[\tilde{H}^{(S)}\slash{n}+\tilde{E}^{(S)}\frac{n\cdot(p'-p)}{2M}\right]\gamma_{5}.\end{equation}

\subsection{Application to the pion-nucleon distributions}

\begin{lyxlist}{00.00.0000}
\item [\label{section_SPT_application}]~
\end{lyxlist}
From the soft-pion theorems for the matrix elements of twist-2 lightcone
operators, equations (\ref{SPT}), (\ref{SPT_vector_isoscalar}),
(\ref{SPT_gluon}), (\ref{SPT5_isovector}), and (\ref{SPT5_isoscalar}),
we arrive at the corresponding results for the $\pi N$ distributions
$H_{i}$ and $\tilde{H}_{i}$ by Fourier transformation with respect
to the light-cone coordinate $\lambda$ and an appropriate decomposition
of the Dirac matrix structure. These results are presented in the
following at the pion threshold. The general (non-threshold) results
are given in appendix \ref{non-threshold_app}.

The soft-pion theorem (\ref{SPT}) leads to\begin{eqnarray}
H_{1\mathrm{th}}^{(+)}(x,\xi,t) & = & -\frac{2M^{2}(1-\xi)}{2M^{2}-t}H^{(V)}(x,\xi,\Delta^{2})+\frac{4M^{2}\xi-t}{2(2M^{2}-t)}E^{(V)}(x,\xi,\Delta^{2})\label{H1thPlus}\\
H_{2\mathrm{th}}^{(+)}(x,\xi,t) & = & \frac{t[H^{(V)}(x,\xi,\Delta^{2})+E^{(V)}(x,\xi,\Delta^{2})]}{2(2M^{2}-t)}\end{eqnarray}
 and\begin{eqnarray}
H_{1\mathrm{th}}^{(-)}(x,\xi,t) & = & \frac{\xi_{0}}{g_{A}}[\tilde{E}^{(V)}(x,\xi_{0},t)-\tilde{E}_{\mathrm{pole}}^{(V)}(x,\xi_{0},t)]\nonumber \\
 &  & +\frac{2M^{2}(1-\xi)}{2M^{2}-t}H^{(V)}(x,\xi,\Delta^{2})-\frac{4M^{2}\xi-t}{2(2M^{2}-t)}E^{(V)}(x,\xi,\Delta^{2})\nonumber \\
 &  & +\frac{2M^{2}}{m_{\pi}^{2}-t}H_{\pi}^{(V)}\left(\frac{x}{\bar{p}_{t}\cdot n},\frac{\xi}{\bar{p}_{t}\cdot n},0\right)\,\theta(\bar{p}_{t}\cdot n-|x|)\\
H_{2\mathrm{th}}^{(-)}(x,\xi,t) & = & -\frac{\tilde{H}(x,\xi_{0},t)}{g_{A}}-\frac{t[H^{(V)}(x,\xi,\Delta^{2})+E^{(V)}(x,\xi,\Delta^{2})]}{2(2M^{2}-t)}.\label{H2thMinus}\end{eqnarray}
Here, $\Delta^{2}$ is considered as function of $t$ according to
the threshold requirement (\ref{momentum_transfer_th}). The variable
$\xi_{0}$ is defined as\begin{equation}
\xi_{0}=\frac{(p-p')\cdot n}{(p+p')\cdot n},\end{equation}
its threshold value is related to $\xi$ via\begin{equation}
\xi_{0}\stackrel{\mathrm{th}}{=}\frac{(2M+m_{\pi})\xi+m_{\pi}}{2M+m_{\pi}(1+\xi)}.\label{xi_0}\end{equation}
The average momentum in the $t$ channel is referred to as $\bar{p}_{t}$,\begin{equation}
\bar{p}_{t}=\frac{p-p'+k}{2}.\end{equation}
At threshold, $\bar{p}_{t}\cdot n$ can be expressed in terms of $\xi$:\begin{equation}
\bar{p}_{t}\cdot n\stackrel{\mathrm{th}}{=}\xi+\frac{m_{\pi}(1-\xi)}{M+m_{\pi}}.\label{pbar_t_n}\end{equation}
 Further, $\tilde{E}_{\mathrm{pole}}^{(V)}$ denotes the pion-pole
contribution to $\tilde{E}^{(V)}$ and is given in terms of the pion
distribution amplitude $\phi_{\pi}$ as \cite{Frankfurt:1998jq,Mankiewicz:1997aa,Frankfurt:1998et,Penttinen:1999th}\begin{equation}
\tilde{E}_{\mathrm{pole}}^{(V)}(x,\xi_{0},t)=\frac{(2M)^{2}g_{A}}{m_{\pi}^{2}-t}\frac{1}{\xi_{0}}\phi_{\pi}\left(\frac{x}{\xi_{0}}\right)\,\theta(\xi_{0}-|x|).\label{ETiled_PionPole}\end{equation}

Next, we turn to the $\pi N$ distributions for the soft-pion theorems
(\ref{SPT_vector_isoscalar}) and (\ref{SPT_gluon}). They read\begin{eqnarray}
H_{1\mathrm{th}}^{(0,G)}(x,\xi,t) & = & -\frac{2M^{2}(1-\xi)}{2M^{2}-t}\, H^{(S,G)}(x,\xi,\Delta^{2})+\frac{4M^{2}\xi-t}{2(2M^{2}-t)}\, E^{(S,G)}(x,\xi,\Delta^{2})\nonumber \\
 &  & +\frac{2M^{2}}{m_{\pi}^{2}-t}\, H_{\pi}^{(S,G)}\left(\frac{x}{\bar{p}_{t}\cdot n},\frac{\xi}{\bar{p}_{t}\cdot n},0\right)\,\theta(\bar{p}_{t}\cdot n-|x|)\\
H_{2\mathrm{th}}^{(0,G)}(x,\xi,t) & = & \frac{t[H^{(S,G)}(x,\xi,\Delta^{2})+E^{(S,G)}(x,\xi,\Delta^{2})]}{2(2M^{2}-t)},\end{eqnarray}
where the first superscript refers to the isoscalar quark distributions
and the second one to the gluon distributions, respectively.

Finally, we give the results related to the soft-pion theorems (\ref{SPT5_isovector})
and (\ref{SPT5_isoscalar}), again using a combined notation for the
isoscalar $(0)$ and isovector even $(+)$ GPDs:\begin{eqnarray}
\tilde{H}_{1\mathrm{th}}^{(0,+)}(x,\xi,t) & = & \frac{2M^{2}(1-\xi)}{2M^{2}-t}\tilde{H}^{(S,V)}(x,\xi,\Delta^{2})-\frac{t\xi\tilde{E}^{(S,V)}(x,\xi,\Delta^{2})}{2(2M^{2}-t)}\label{H1thPlusTilde}\\
\tilde{H}_{2\mathrm{th}}^{(0,+)}(x,\xi,t) & = & -\frac{4M^{2}-t}{2(2M^{2}-t)}\,\tilde{H}^{(S,V)}(x,\xi,\Delta^{2})\end{eqnarray}
and\begin{eqnarray}
\tilde{H}_{1\mathrm{th}}^{(-)}(x,\xi,t) & = & \frac{E^{(V)}(x,\xi_{0},t)}{g_{A}}-\frac{2M^{2}(1-\xi)}{2M^{2}-t}\tilde{H}^{(V)}(x,\xi,\Delta^{2})\nonumber \\
 &  & +\frac{t\xi\tilde{E}^{(V)}(x,\xi,\Delta^{2})}{2(2M^{2}-t)}-\frac{2Mm_{\pi}}{g_{A}t}\phi_{\pi}\left(\frac{x}{\xi}\right)\,\theta(\xi-|x|)\\
\tilde{H}_{2\mathrm{th}}^{(-)}(x,\xi,t) & = & \frac{4M^{2}-t}{2(2M^{2}-t)}\tilde{H}^{(V)}(x,\xi,\Delta^{2})-\frac{E^{(V)}(x,\xi_{0},t)+H^{(V)}(x,\xi_{0},t)}{g_{A}}.\label{H2thMinusTilde}\end{eqnarray}

\subsection{Results for the moments of $\pi N$ distributions}

In section \ref{section_moments}, we have described how to obtain
pion emission form factors of certain local operators from the moments
of the $\pi N$ distributions according to the polynomiality property.
Now that we have derived the soft-pion theorems for the $\pi N$ distributions,
we can easily read off these form factors after taking the corresponding
moment. In the cases where the results are known, this procedure provides
a check of the previous calculations.

\subsubsection{First moment of $H_{i}^{(0,\pm)}$}

The soft-pion theorems for pion emission induced by the local vector
current, i.e.~for the form factors $A_{i}$ ($i=1,\ldots,8$) that
we have defined in formula (\ref{electroproduction_amplitude}), are
given in appendix \ref{appendix_local_vector}. For a discussion,
we restrict ourselves in the following to the pion threshold where
the number of independent form factors is reduced to two. The conventional
quantities for such a description are the transversal and longitudinal
$s$ wave multipoles $E^{(0,\pm)}$ and $L^{(0,\pm)}$. In the center-of-mass
frame where $\vec{p}\,'=\vec{k}=0$ and $\vec{p}=-\vec{\Delta}$,
these threshold multipoles can be defined using the spatial components
of the matrix elements\[
\langle N_{f}(p',S')\,\pi^{a}(k)|\bar{\psi}\gamma_{\mu}\left\{ \begin{array}{c}
1\\
\tau^{b}\end{array}\right\} \psi|N_{i}(p,S)\rangle=\i\,\left\{ \begin{array}{c}
T_{S'S}^{\mu(0)}\tau_{fi}^{a}\\
T_{S'S}^{\mu(+)}\delta^{ab}\delta_{fi}+T_{S'S}^{\mu(-)}\i\varepsilon^{abc}\tau_{fi}^{c}\end{array}\right\} ,\]
namely\begin{eqnarray}
\frac{e}{6}\vec{T}_{S'S}^{(0)} & = & 8\pi(M+m_{\pi})\left[E_{0+}^{(0)}\vec{\sigma}_{S'S}+(L_{0+}^{(0)}-E_{0+}^{(0)})\frac{\vec{p}\,\vec{p}\cdot\vec{\sigma}_{S'S}}{|\vec{p}|^{2}}\right]\\
\frac{e}{2}\vec{T}_{S'S}^{(\pm)} & = & 8\pi(M+m_{\pi})\left[E_{0+}^{(\pm)}\vec{\sigma}_{S'S}+(L_{0+}^{(\pm)}-E_{0+}^{(\pm)})\frac{\vec{p}\,\vec{p}\cdot\vec{\sigma}_{S'S}}{|\vec{p}|^{2}}\right],\end{eqnarray}
where $e$ is the electromagnetic coupling constant and $\vec{\sigma}=(\sigma_{x},\sigma_{y},\sigma_{z})$
are the Pauli spin matrices.

\begin{figure}[!t]
\begin{center}\includegraphics[%
  width=0.40\textwidth]{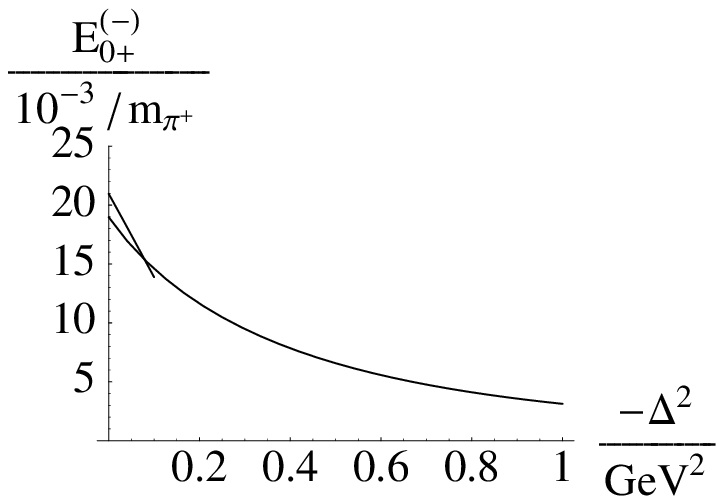}\includegraphics[%
  width=0.40\textwidth]{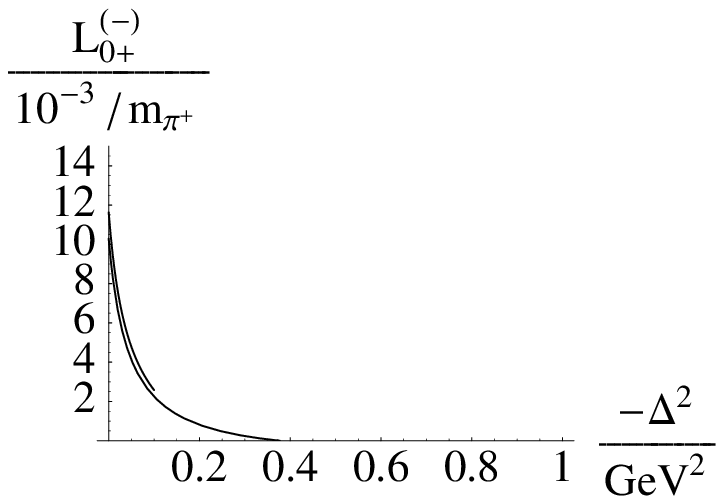}\end{center}

\begin{center}\includegraphics[%
  width=0.40\textwidth]{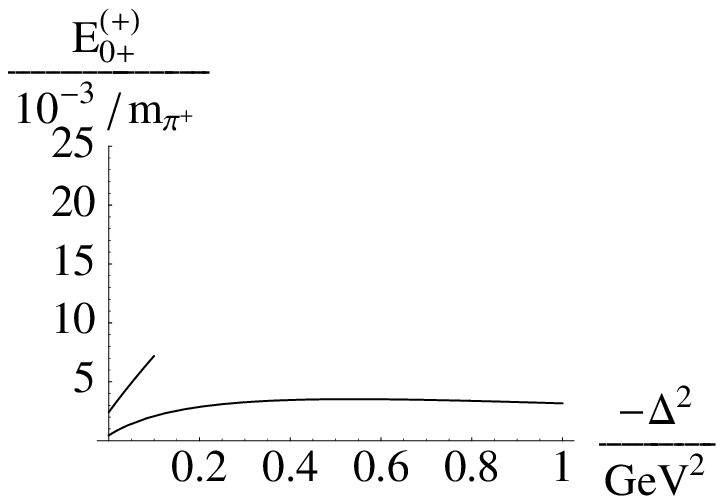}\includegraphics[%
  width=0.40\textwidth]{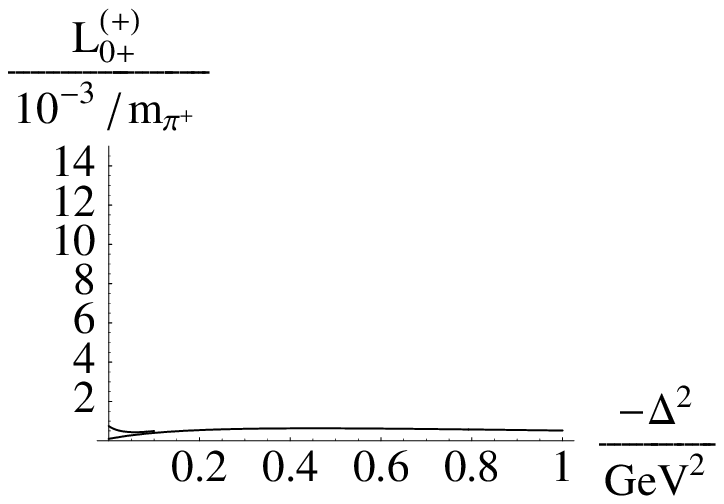}\end{center}

\begin{center}\includegraphics[%
  width=0.40\textwidth]{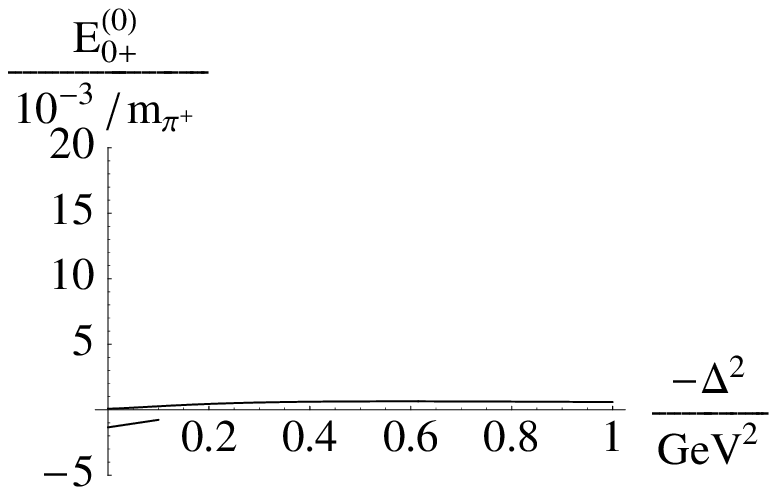}\includegraphics[%
  width=0.40\textwidth]{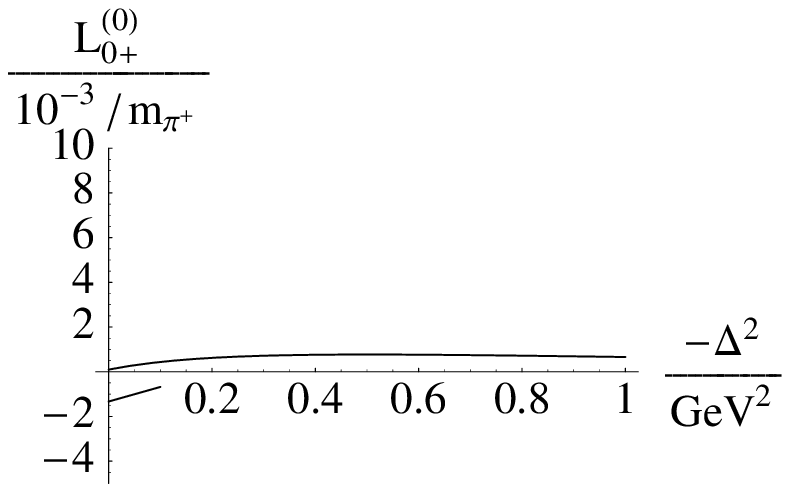}\end{center}

\caption{\label{figure_multipoles}The transverse and longitudinal $s$-wave
multipoles as function of $\Delta^{2}=(1+m_{\pi}/M)t+m_{\pi}^{2}$.
The curves up to $-\Delta^{2}=1\textrm{ GeV}^{2}$ correspond to the
soft-pion theorems (\ref{E0+^0}) to (\ref{L0+^-}). The curves up
to $-\Delta^{2}=0.1\textrm{ GeV}^{2}$ show the one-loop ChPT results
by Bernard et.~al.~\cite{Bernard:1994bq} as given in their formula
(5.1). The references for the nucleon form factors which have been
used for this plot can be found in appendix \ref{GPDs_app}. For the
pion electromagnetic form factor we have used the monopole form $F_{\pi}(\Delta^{2})=(1-\Delta^{2}\langle r_{\pi}^{2}\rangle/6)^{-1}$
with $\langle r_{\pi}^{2}\rangle\approx0.45\textrm{ fm}^{2}$, see
references \cite{Volmer:2000ek,Bijnens:2002hp}.}\lyxline{\normalsize}

\end{figure}

According to these definitions, we obtain for the relations between
these multipoles and the $\pi N$ distributions at threshold\begin{eqnarray}
E_{0+}^{(0)} & = & -\frac{ec}{3}\int\limits _{-1}^{1}\d x\, H_{2\mathrm{th}}^{(0)}(x,\xi,t)=\frac{ec}{3}\left[-\frac{tG_{M}^{(S)}(\Delta^{2})}{2(2M^{2}-t)}+\mathcal{O}(\varepsilon)\right]\label{E0+^0}\\
E_{0+}^{(+)} & = & -ec\int\limits _{-1}^{1}\d x\, H_{2\mathrm{th}}^{(+)}(x,\xi,t)=ec\left[-\frac{tG_{M}^{(V)}(\Delta^{2})}{2(2M^{2}-t)}+\mathcal{O}(\varepsilon)\right]\\
E_{0+}^{(-)} & = & -ec\int\limits _{-1}^{1}\d x\, H_{2\mathrm{th}}^{(-)}(x,\xi,t)=ec\left[F_{A}^{(V)}(t)+\frac{tG_{M}^{(V)}(\Delta^{2})}{2(2M^{2}-t)}+\mathcal{O}(\varepsilon)\right],\end{eqnarray}
and\begin{eqnarray}
L_{0+}^{(0)} & = & E_{0+}^{(0)}+\frac{ect}{12M^{2}}\int\limits _{-1}^{1}\d x\, H_{1\mathrm{th}}^{(0)}(x,1,t)=\frac{ec}{3}\left[-\frac{tG_{E}^{(S)}(\Delta^{2})}{2(2M^{2}-t)}+\mathcal{O}(\varepsilon)\right]\\
L_{0+}^{(+)} & = & E_{0+}^{(\pm)}+\frac{ect}{4M^{2}}\int\limits _{-1}^{1}\d x\, H_{1\mathrm{th}}^{(+)}(x,1,t)=ec\left[-\frac{tG_{E}^{(V)}(\Delta^{2})}{2(2M^{2}-t)}+\mathcal{O}(\varepsilon)\right]\\
L_{0+}^{(-)} & = & E_{0+}^{(-)}+\frac{ect}{4M^{2}}\int\limits _{-1}^{1}\d x\, H_{1\mathrm{th}}^{(-)}(x,1,t)=ec\left[\frac{m_{\pi}^{2}F_{A}^{(V)}(t)}{m_{\pi}^{2}-t}+\frac{tG_{E}^{(V)}(\Delta^{2})}{2(2M^{2}-t)}+\mathcal{O}(\varepsilon)\right],\quad\label{L0+^-}\end{eqnarray}
with the kinematical prefactor\begin{equation}
c=\frac{g_{A}}{16\pi f_{\pi}}\frac{\sqrt{4M^{2}-t}}{M+m_{\pi}}.\label{c}\end{equation}
The functions \begin{equation}
G_{M}(\Delta^{2})=F_{1}(\Delta^{2})+F_{2}(\Delta^{2}),\qquad G_{E}(\Delta^{2})=F_{1}(\Delta^{2})+\frac{\Delta^{2}}{4M^{2}}F_{2}(\Delta^{2})\end{equation}
 are the magnetic and electric form factors of the nucleon, with the
superscripts $S$ and $V$ indicating the isovector and isoscalar
combination, respectively. $F_{A}^{(V)}\equiv G_{A}^{(V)}/g_{A}$
denotes the isovector axial form factor normalized to unity. Figure
\ref{figure_multipoles} shows a plot of the multipoles $E_{0+}$
and $L_{0+}$ as function of the invariant momentum transfer $\Delta^{2}$.%

The multipoles $E_{0+}^{(0,+)}$ and $L_{0+}^{(0,+)}$ turn out to
be quite small, so that the corrections which are neglected in the
soft-pion theorems become important for these quantities. In the conventional
units that are also used in figure \ref{figure_multipoles}, these
corrections could be as large as $cm_{\pi}/M\sim3\times10^{-3}/m_{\pi^{+}}$.
The situation is different for the multipoles $E_{0+}^{(-)}$ and
$L_{0+}^{(-)}$; they reach sizable values in particular at low momentum
transfer, and corrections become relatively small.

Since the first use of soft-pion theorems in pion-electroproduction
that can be dated back to the work of Nambu and Shrauner in 1962 \cite{Nambu:1962wb},
a lot of progress has been made in calculating corrections to the
early soft-pion results. Within the framework of chiral perturbation
theory, pion-electroproduction has been computed including the one-loop
level \cite{Bernard:1994bq}. The resulting threshold multipoles are
also shown in figure \ref{figure_multipoles}. Unfortunately, these
achievements of ChPT are limited to the region of very small momentum
transfer, $-t\ll M^{2}$. If we expand the soft-pion results (\ref{E0+^0})
to (\ref{L0+^-}) for such small $-t$, we get correspondence with
the leading-order results of ChPT.

\subsubsection{First moment of $\tilde{H}_{i}^{(\pm)}$}

Further, we would like to discuss the form factors of pion emission
induced by the local isovector axial current. They can be obtained
from the first moments of the $\pi N$ distributions $\tilde{H}_{i}^{(\pm)}$.
Again, we restrict ourselves to the threshold. Then the matrix element
is parametrized in terms of three $s$-wave multipoles $L_{0+}^{(\pm)}$,
$H_{0+}^{(\pm)}$, and $M_{0+}^{(\pm)}$:\begin{equation}
\langle N_{f}(p',S')\,\pi^{a}(k)|A^{\mu a}|N_{i}(p,S)\rangle=\frac{\i}{2}(\delta^{ab}\delta_{fi}\tilde{T}_{S'S}^{\mu(+)}+\i\varepsilon^{abc}\tau_{fi}^{c}\tilde{T}_{S'S}^{\mu(-)}),\end{equation}
where in the center-of-mass frame ($\vec{p}\,'+\vec{k}=0$) one defines
for the time and space components of $\tilde{T}^{\mu}$ \begin{eqnarray}
\tilde{T}_{S'S}^{0(\pm)} & = & 16\pi(M+m_{\pi})(L_{0+}^{(\pm)}+\Delta_{0}H_{0+}^{(\pm)})\delta_{S'S},\\
\vec{\tilde{T}}_{S'S}^{(\pm)} & = & 16\pi(M+m_{\pi})\left(-\vec{p}H_{0+}^{(\pm)}\delta_{S'S}+M_{0+}^{(\pm)}\i\frac{\vec{\sigma}_{S'S}\times\vec{p}}{|\vec{p}|}\right)\end{eqnarray}
(see \cite{Bernard:1994xh} and original references therein). The
relations between these multipoles and the moments of theshold $\pi N$
distributions are\begin{eqnarray}
L_{0+}^{(\pm)} & = & c\int\limits _{-1}^{1}\d x\left(\frac{M+m_{\pi}}{M}\tilde{H}_{1\mathrm{th}}^{(\pm)}(x,0,t)+\frac{2M(2M+m_{\pi})}{4M^{2}-t}\tilde{H}_{2\mathrm{th}}^{(\pm)}(x,\xi,t)\right)\\
H_{0+}^{(\pm)} & = & \frac{c}{M}\int\limits _{-1}^{1}\d x\left(-\frac{1}{2}\tilde{H}_{1\mathrm{th}}^{(\pm)}(x,1,t)-\frac{2M^{2}}{4M^{2}-t}\tilde{H}_{2\mathrm{th}}^{(\pm)}(x,\xi,t)\right)\\
M_{0+}^{(\pm)} & = & -\frac{g_{A}}{16\pi f_{\pi}}\frac{\sqrt{-t}}{M+m_{\pi}}\int\limits _{-1}^{1}\d x\,\tilde{H}_{2}^{(\pm)}(x,\xi,t),\end{eqnarray}
where $c$ is the same kinematical factor as in definition (\ref{c}).
Note that in taking the moments of the $\pi N$ distributions at a
particular skewedness $\xi$, in general one has to integrate first
and then to insert the value of $\xi$ to avoid unphysical regions
in the $\xi$-$t$ plane. After the insertion of the soft-pion theorems
for the functions $\tilde{H}_{i\mathrm{th}}$, we obtain\begin{eqnarray}
M_{0+}^{(-)} & = & \frac{g_{A}}{16\pi f_{\pi}}\sqrt{-\frac{t}{M^{2}}}\left[\frac{G_{M}^{(V)}(t)}{g_{A}}-\frac{(4M^{2}-t)G_{A}^{(V)}(t)}{2(2M^{2}-t)}\right]\\
M_{0+}^{(+)} & = & \frac{g_{A}}{16\pi f_{\pi}}\sqrt{-\frac{t}{M^{2}}}\frac{(4M^{2}-t)G_{A}^{(V)}(t)}{2(2M^{2}-t)}\end{eqnarray}
and\begin{equation}
L_{0+}^{(-)}=-c\frac{4MG_{E}^{(V)}(t)}{4M^{2}-t},\qquad H_{0+}^{(-)}=\frac{c}{M}\left[\frac{m_{\pi}/g_{A}}{\Delta^{2}-m_{\pi}^{2}}+\frac{2MG_{E}^{(V)}(t)}{(4M^{2}-t)g_{A}}\right],\end{equation}
while $L_{0+}^{(+)}$ and $H_{0+}^{(+)}$ are vanishing within the
accuracy of soft-pion theorems. At small momentum transfer, these
results agree with the leading order expressions of the ChPT calculation
in reference \cite{Bernard:1994xh}.

\subsubsection{Second moment of $H_{i}^{(0)}+H_{i}^{(G)}/2$}

As described in section \ref{section_moments}, the second moment
of the $\pi N$ distributions $H_{i}^{(0)}$ and $H_{i}^{(G)}$ give
the form factors $B_{i}$ for soft pion emission induced by the energy-momentum
tensor. The resulting soft-pion theorem is\begin{eqnarray}
\lefteqn{\langle N'\pi^{a}|\mathcal{T}^{\mu\nu}|N\rangle=\frac{\i g_{A}}{Mf_{\pi}}\sum_{i=1}^{20}\bar{U}'B_{i}\Gamma_{i}^{\mu\nu}\tau^{a}U}\nonumber \\
 & = & \frac{\i g_{A}}{Mf_{\pi}}\bar{U}'\left\{ \left[\Delta^{2}\left(C-\frac{M^{2}}{m_{\pi}^{2}-t}\right)-M^{2}\right]\frac{g^{\mu\nu}}{4}+B\bar{p}^{\mu}\bar{p}^{\nu}-\frac{C}{4}\Delta^{\mu}\Delta^{\nu}+\frac{M^{2}k^{\mu}k^{\nu}}{4(m_{\pi}^{2}-t)}\right.\nonumber \\
 &  & +\frac{2M^{2}(A+B)}{u-M^{2}}\bar{p}^{\{\mu}k^{\nu\}}-\frac{M^{2}\Delta^{\{\mu}k^{\nu\}}}{m_{\pi}^{2}-t}-\left(\frac{M}{W^{2}-M^{2}}+\frac{M}{u-M^{2}}\right)\slash k\nonumber \\
 &  & \left.\times\left[\frac{\Delta^{2}}{4}Cg^{\mu\nu}+B\bar{p}^{\mu}\bar{p}^{\nu}-\frac{C}{4}\Delta^{\mu}\Delta^{\nu}+(A+B)\bar{p}^{\{\mu}\gamma^{\nu\}}\right]\right\} \gamma_{5}\tau^{a}U+\mathcal{O}(\varepsilon),\end{eqnarray}
where $A=A(\Delta^{2})$ and so on for the other nucleon form factors
$B$ and $C$. For the determination of $B_{1}$ and $B_{8}$ from
the $\pi N$ distribution moments, it has been necessary to use the
current conservation relations (\ref{T_current_conservation1}) and
(\ref{T_current_conservation_4}), because these form factors appear
together with the metric tensor $g^{\mu\nu}$ which vanishes after
contraction with the lightcone vectors $n^{\mu}n^{\nu}$. The remaining
six current conservation constraints are fulfilled, so that\begin{equation}
\Delta_{\mu}\langle N'\pi^{a}|\mathcal{T}^{\mu\nu}|N\rangle=\mathcal{O}(\varepsilon).\end{equation}

\section{Hard pion production with additional soft pion}

As an application of the presented soft pion theorems for the $\pi N$
distributions, we consider the process of hard $\pi^{+}$ production
off the proton with soft pion emission. The two possible reactions
are\begin{equation}
\gamma_{L}^{*}(q)+\textrm{p}(p,S)\rightarrow\textrm{n}(p',S')+\pi^{0}(k)+\pi^{+}(q'),\label{HEMP+pi0}\end{equation}
and\begin{equation}
\gamma_{L}^{*}(q)+\textrm{p}(p,S)\rightarrow\textrm{p}(p',S')+\pi^{-}(k)+\pi^{+}(q').\label{HEMP+pim}\end{equation}
We shall calculate the longitudinal differential cross section as
well as the transverse spin asymmetries for these processes, in comparison
to the corresponding pure process, namely \begin{equation}
\gamma_{L}^{*}(q)+\textrm{p}(p,S)\rightarrow\textrm{n}(p',S')+\pi^{+}(q').\label{HEMP}\end{equation}

\subsection{Longitudinal cross section}

First, we recall the definition of the differential longitudinal cross
section of $\pi^{+}$ production without soft pion emission, reaction
(\ref{HEMP}):\begin{equation}
\d^{2}\sigma_{L}^{(\mathrm{n})}=\frac{\overline{|\M_{L}^{(\mathrm{n})}|^{2}}\,\d^{2}\Phi(q+p;q',p')}{2(s-M^{2})\sqrt{\Lambda(s,M^{2},-Q^{2})}}\label{dsig}\end{equation}
where $\d^{2}\Phi$ is the two-particle phase space volume,\begin{equation}
\d^{2}\Phi(q+p;q',p')=\frac{\d^{3}q'}{2q_{0}'}\frac{\d^{3}p'}{2p_{0}'}(2\pi)^{4}\delta(q+p-q'-p'),\label{d2Phi}\end{equation}
and $\Lambda$ denotes the conventional kinematical function \begin{equation}
\Lambda(x,y,z)=\sqrt{(x-y-z)^{2}-4yz}.\end{equation}
Introducing the angle $\phi$, which is the azimuthal angle of $\vec{q}\,'$
with respect to the direction of $\vec{q}$ in the center-of-mass
frame, and $t_{0}=(p-p')^{2}$, one obtains for the cross section\begin{equation}
\frac{\d^{2}\sigma_{L}^{(\mathrm{n})}}{\d\phi\,\d t_{0}}=\frac{\overline{|\M_{L}^{(\mathrm{n})}|^{2}}}{32\pi^{2}(s-M^{2})\Lambda(s,M^{2},-Q^{2})}.\end{equation}
The amplitude is given through the matrix element\begin{equation}
\M_{L}^{(\mathrm{n})}=e\langle\textrm{n}(p')\,\pi^{+}(q')|\varepsilon_{L}\cdot J|\textrm{p}(p)\rangle\end{equation}
of the electromagnetic current\begin{equation}
J=\bar{\psi}\gamma_{\mu}\left(\frac{1}{6}+\frac{\tau^{3}}{2}\right)\psi,\end{equation}
where $\varepsilon_{L}$ is the longitudinal polarization vector,
defined in the center-of-mass frame to be\begin{equation}
\varepsilon_{L}=\frac{1}{Q}\left(|\vec{q}|,q_{0}\frac{\vec{q}}{|\vec{q}|}\right).\end{equation}

In the case of the reactions (\ref{HEMP+pi0}) and (\ref{HEMP+pim}),
i.e.~with soft pion emission, we have to replace (\ref{d2Phi}) with
the three-particle phase space\begin{equation}
\d^{5}\Phi(q+p;q',p',k)=\frac{\d^{3}q'}{2q_{0}'}\frac{\d^{3}p'}{2p_{0}'}\frac{\d^{3}k}{2k_{0}}(2\pi)^{4}\delta(q+p+k-q'-p'),\end{equation}
so that we obtain a differential cross section \begin{equation}
\d^{5}\sigma_{L}^{(N\pi)}=\frac{\overline{|\M_{L}^{(N\pi)}|^{2}}\,\d^{5}\Phi(q+p;q',p',k)}{2(s-M^{2})\sqrt{\Lambda(s,M^{2},-Q^{2})}}\end{equation}
 defined analogously to equation (\ref{dsig}). The amplitude is \begin{equation}
\M_{L}^{(N\pi)}=e\langle N(p')\pi(k)\,\pi^{+}(q')|\varepsilon_{L}\cdot J|\textrm{p}(p)\rangle,\end{equation}
and the superscript $N\pi$ labels the two final state possibilities
$\textrm{n}\pi^{0}$ or $\textrm{p}\pi^{-}$. Integrating out the
angular dependence of the soft pion and the invariant mass $W^{2}$
of the final nucleon-pion system up to some (not too large) value
$W_{\mathrm{max}}^{2}$, we further define\begin{equation}
\frac{\d^{2}\sigma_{L}^{(N\pi)}}{\d\phi\,\d\Delta^{2}}\equiv\int\limits _{W_{\mathrm{th}}^{2}}^{W_{\mathrm{max}}^{2}}\d W^{2}\int\d\Omega_{\pi}\frac{\d^{5}\sigma_{L}^{(N\pi)}}{\d\phi\,\d\Delta^{2}\,\d W^{2}\d\Omega_{\pi}}.\label{dsigdphidDelta}\end{equation}
Here, the variable $\Omega_{\pi}$ denotes the solid angle of the
soft pion in the center-of-mass frame of the final nucleon-pion system.
If the soft pion momentum $k$ is sufficiently small, we can approximate
the amplitude $\M_{L}^{(N\pi)}$ by its threshold value, and perform
the phase-space integration in (\ref{dsigdphidDelta}) to obtain\begin{equation}
\frac{\d^{2}\sigma_{L}^{(N\pi)}}{\d\phi\,\d\Delta^{2}}=\frac{\overline{|\M_{L}^{(N\pi)}|^{2}}\,\Phi(W_{\mathrm{max}})}{32\pi^{2}(s-M^{2})\Lambda(s,M^{2},-Q^{2})},\end{equation}
with the phase-space function\begin{equation}
\Phi(W_{\mathrm{max}})=\frac{1}{6\pi^{2}}\sqrt{\frac{2Mm_{\pi}}{M+m_{\pi}}}(W_{\mathrm{max}}-W_{\mathrm{th}})^{3/2}\left[1+\mathcal{O}\left(\frac{W_{\mathrm{max}}}{W_{\mathrm{th}}}-1\right)\right].\label{Phi}\end{equation}

\subsection{Transverse spin asymmetry}

Besides the longitudinal cross section itself, there exist predictions
for the so-called transverse spin asymmetry in $\pi^{+}$ production
(\ref{HEMP}) with a polarized target proton \cite{Frankfurt:1999fp,Frankfurt:1999xe}.
There were arguments brought forward that this observable is particularly
useful, because it is less sensitive to higher-twist corrections and
next-to-leading order corrections in the strong coupling $\alpha_{s}$
\cite{Belitsky:2001nq}. The definition of this asymmetry
is\begin{equation}
\mathcal{A}_{\mathrm{n}}=\frac{1}{|\vec{S}_{\bot}|}\left(\int\limits _{0}^{\pi}\d\phi\frac{\d^{2}\sigma_{L}^{(\mathrm{n})}}{\d\phi\,\d t_{0}}-\int\limits _{\pi}^{2\pi}\d\phi\frac{\d^{2}\sigma_{L}^{(\mathrm{n})}}{\d\phi\,\d t_{0}}\right)\left(\int\limits _{0}^{2\pi}\d\phi\frac{\d^{2}\sigma_{L}^{(\mathrm{n})}}{\d\phi\,\d t_{0}}\right)^{-1},\end{equation}
where $\vec{S}_{\bot}$ is the component of the proton's spin vector
that is transverse to $\vec{q}$ in the center-of-mass frame. A splitting
of the squared amplitude into spin-dependent and spin-independent
parts yields\begin{equation}
\sum_{S'}|\M_{L}^{(\mathrm{n})}|^{2}\propto s_{0}^{(\mathrm{n})}(x_{B},t_{0})+s_{1}^{(\mathrm{n})}(x_{B},t_{0})|\vec{S}_{\bot}|\sin\phi,\label{sumMn}\end{equation}
so that one finds\begin{equation}
\mathcal{A}_{\mathrm{n}}=\frac{2s_{1}^{(\mathrm{n})}}{\pi s_{0}^{(\mathrm{n})}}.\end{equation}

It is now straightforward to define an appropriate generalization
of this observable for the equivalent process with soft pion emission:\begin{equation}
\mathcal{A}_{N\pi}=\frac{1}{|\vec{S}_{\bot}|}\left(\int\limits _{0}^{\pi}\d\phi\frac{\d^{2}\sigma_{L}^{(N\pi)}}{\d\phi\,\d\Delta^{2}}-\int\limits _{\pi}^{2\pi}\d\phi\frac{\d^{2}\sigma_{L}^{(N\pi)}}{\d\phi\,\d\Delta^{2}}\right)\left(\int\limits _{0}^{2\pi}\d\phi\frac{\d^{2}\sigma_{L}^{(N\pi)}}{\d\phi\,\d\Delta^{2}}\right)^{-1}=\frac{2s_{1}^{(N\pi)}}{\pi s_{0}^{(N\pi)}},\end{equation}
where the functions $s_{0}^{(N\pi)}$ and $s_{1}^{(N\pi)}$ arise
from the following decomposition of the threshold amplitude:\begin{equation}
\sum_{S'}|\M_{L}^{(N\pi)}|^{2}\propto\frac{1}{f_{\pi}^{2}}[s_{0}^{(N\pi)}(x_{B},\Delta^{2})+s_{1}^{(N\pi)}(x_{B},\Delta^{2})|\vec{S}_{\bot}|\sin\phi].\label{Sum_MNpi}\end{equation}
The constant of proportionality not written explicitely equals that
one in (\ref{sumMn}), it contains the distribution amplitude of the
$\pi^{+}$ and the strong coupling $\alpha_{s}$, for example. Supplementing
the factor $f_{\pi}^{2}$ in the denominator serves to keep the functions
$s_{0}^{(N\pi)}$ and $s_{1}^{(N\pi)}$ dimensionless.

Of course, from an experimental point of view, these asymmetries are
useful only when the detection of the soft pion is guaranteed. If
this is not the case, one should formulate an asymmetry with the cross
sections of all three processes, (\ref{HEMP+pi0}) to (\ref{HEMP}),
added up. In terms of the functions $s_{0}$ and $s_{1}$, such an
asymmetry reads\begin{equation}
\mathcal{A}_{\mathrm{n}+\mathrm{n}\pi^{0}+\mathrm{p}\pi^{-}}=\frac{2}{\pi}\frac{s_{1}^{(\mathrm{n})}+(s_{1}^{(\mathrm{n}\pi^{0})}+s_{1}^{(\mathrm{p}\pi^{-})})\Phi(W_{\mathrm{max}})/f_{\pi}^{2}}{s_{0}^{(\mathrm{n})}+(s_{0}^{(\mathrm{n}\pi^{0})}+s_{0}^{(\mathrm{p}\pi^{-})})\Phi(W_{\mathrm{max}})/f_{\pi}^{2}}.\label{A_n+npi0+ppi-}\end{equation}

\subsection{Amplitude at threshold}

Now we come to the amplitude $\M_{L}^{(N\pi)}$ of hard pion production
with soft pion emission. For the sake of generality, we give the amplitude
for arbitrary pion isospins $a$ and $c$:\begin{eqnarray}
\lefteqn{e\langle N(p')\pi^{a}(k)\,\pi^{c}(q')|\varepsilon_{L}\cdot J|N(p)\rangle}\nonumber \\
 & = & -\i e\frac{2}{9}\frac{4\pi\alpha_{s}}{Q}\int\limits _{-1}^{1}\d u\frac{f_{\pi}\phi_{\pi}(u)}{1+u}\int\limits _{-1}^{1}\d x\frac{\i g_{A}}{Mf_{\pi}}\bar{U}'[c^{+}\delta^{3c}\tau^{a}(\tilde{H}_{1\mathrm{th}}^{(0)}+\tilde{H}_{2\mathrm{th}}^{(0)}M\slash{n})\nonumber \\
 &  & +(c^{+}\delta^{ac}/3+c^{-}\i\varepsilon^{c3a})(\tilde{H}_{1\mathrm{th}}^{(+)}+\tilde{H}_{2\mathrm{th}}^{(+)}M\hat{n})\nonumber \\
 &  & +(c^{+}\i\varepsilon^{acd}\tau^{d}/3+c^{-}(\delta^{ac}\tau^{3}-\delta^{a3}\tau^{c})(\tilde{H}_{1\mathrm{th}}^{(-)}+\tilde{H}_{2\mathrm{th}}^{(-)}M\hat{n})]\gamma_{5}U,\end{eqnarray}
where we abbreviate\begin{equation}
c^{\pm}=\frac{1}{x-\xi+\i0}\pm\frac{1}{x+\xi-\i0}.\end{equation}
Evaluating this expression for the isospin combinations needed in
$\pi^{+}$ production, and writing the amplitudes with the help of
two functions $A_{N\pi}$ and $C_{N\pi}$ in the form\begin{equation}
\M_{L}^{(N\pi)}=\frac{4\sqrt{2}e}{9}\frac{4\pi\alpha_{s}}{Q}\int\limits _{-1}^{1}\d u\frac{\phi_{\pi}(u)}{1+u}\,\bar{u}(p')\left(A_{N\pi}\hat{n}+\frac{C_{N\pi}}{M}\right)\gamma_{5}u(p),\end{equation}
we obtain\begin{eqnarray}
A_{\mathrm{n}\pi^{0}} & = & -g_{A}\int\limits _{-1}^{1}\d x\left(\frac{2/3}{x-\xi+\i0}-\frac{1/3}{x+\xi-\i0}\right)\tilde{H}_{2\mathrm{th}}^{(-)}(x,\xi,t)\\
C_{\mathrm{n}\pi^{0}} & = & -g_{A}\int\limits _{-1}^{1}\d x\left(\frac{2/3}{x-\xi+\i0}-\frac{1/3}{x+\xi-\i0}\right)\tilde{H}_{1\mathrm{th}}^{(-)}(x,\xi,t)\end{eqnarray}
for the $\textrm{n}\pi^{0}$ final state, and\begin{eqnarray}
A_{\mathrm{p}\pi^{-}} & = & \frac{g_{A}}{\sqrt{2}}\int\limits _{-1}^{1}\d x\left(\frac{2/3}{x-\xi+\i0}-\frac{1/3}{x+\xi-\i0}\right)[\tilde{H}_{2\mathrm{th}}^{(+)}(x,\xi,t)+\tilde{H}_{2\mathrm{th}}^{(-)}(x,\xi,t)]\\
C_{\mathrm{p}\pi^{-}} & = & \frac{g_{A}}{\sqrt{2}}\int\limits _{-1}^{1}\d x\left(\frac{2/3}{x-\xi+\i0}-\frac{1/3}{x+\xi-\i0}\right)[\tilde{H}_{1\mathrm{th}}^{(+)}(x,\xi,t)+\tilde{H}_{1\mathrm{th}}^{(-)}(x,\xi,t)]\end{eqnarray}
for the $\textrm{p}\pi^{-}$ final state.

Squaring the amplitude and summing over the final nucleon spin $S'$,
we obtain the functions $s_{0}^{(N\pi)}$ and $s_{1}^{(N\pi)}$ that
we have introduced within the decomposition of the squared amplitude
(\ref{Sum_MNpi}):\begin{eqnarray}
s_{0}^{(N\pi)} & = & 4[(\textrm{Re}A_{N\pi})^{2}+(\textrm{Im}A_{N\pi})^{2}]\frac{M(1-\xi^{2})}{M+m_{\pi}}\nonumber \\
 &  & +4\frac{2\xi+m_{\pi}(1+\xi)}{M+m_{\pi}}\textrm{Re}(A_{N\pi}C_{N\pi}^{*})-\frac{t}{M^{2}}[(\textrm{Re}C_{N\pi})^{2}+(\textrm{Im}C_{N\pi})^{2}]\\
s_{1}^{(N\pi)} & = & -4\frac{\sqrt{-\Delta_{\bot}^{2}}}{M+m_{\pi}}\textrm{Im}(A_{N\pi}C_{N\pi}^{*}).\end{eqnarray}
Note that when setting the pion mass equal to zero everywhere in the
last expression, we recover the known formula for the reaction without
soft pion emission.

\subsection{Numerical results}

\subsubsection{Ratio of the longitudinal cross sections}

The longitudinal cross section for usual hard $\pi^{+}$ production
to leading twist and leading order in QCD was given in references
\cite{Mankiewicz:1998kg,Frankfurt:1999fp}. Here, we present the ratios
of the unpolarized cross sections\begin{equation}
\frac{\d\sigma_{L}^{(N\pi)}}{\d\Delta^{2}}\equiv\int\limits _{0}^{2\pi}\d\phi\frac{\d^{2}\sigma_{L}^{(N\pi)}}{\d\phi\,\d\Delta^{2}}\end{equation}
and their corresponding counterpart without soft pion emission, respectively.
These ratios must be calculated for a fixed direction of $\vec{q}\,'$,
which implies the following relation between two- and three-particle
final state variables $\Delta^{2}$ and $t_{0}$: \begin{equation}
\Delta^{2}=-\frac{x_{B}(W_{\mathrm{th}}^{2}-M^{2}+t_{0})-t_{0}}{1-x_{B}}+\mathcal{O}\left(\frac{1}{Q^{2}}\right).\end{equation}
With this specification of $\Delta^{2},$ we can define the ratios
as functions of $x_{B}$ and $t_{0}$:\begin{equation}
R_{N\pi}(x_{B},t_{0},W_{\mathrm{max}})=\frac{\d\sigma_{L}^{(N\pi)}}{\d\Delta^{2}}\left(\frac{\d\sigma_{L}^{(\mathrm{n})}}{\d t_{0}}\right)^{-1}=\frac{s_{0}^{(N\pi)}(x_{B},\Delta^{2})}{s_{0}^{(\mathrm{n})}(x_{B},t_{0})}\frac{\Phi(W_{\mathrm{max}})}{f_{\pi}^{2}}.\label{RNpin}\end{equation}
Concerning the models for the GPDs that enter the calculations of
$s_{0}^{(N\pi)}$ and $s_{0}^{(\mathrm{n})}$, we refer to the review
\cite{Goeke:2001tz}; relevant formulas are summarized in appendix
\ref{GPDs_app}. For the pion distribution amplitude, throughout we
use the asymptotic form \begin{equation}
\phi_{\pi}(u)=\phi_{\mathrm{as}}(u)=\frac{3}{4}(1-u^{2}).\label{phi_as}\end{equation}

\begin{figure}[!t]
\begin{center}\includegraphics[%
  width=0.55\textwidth]{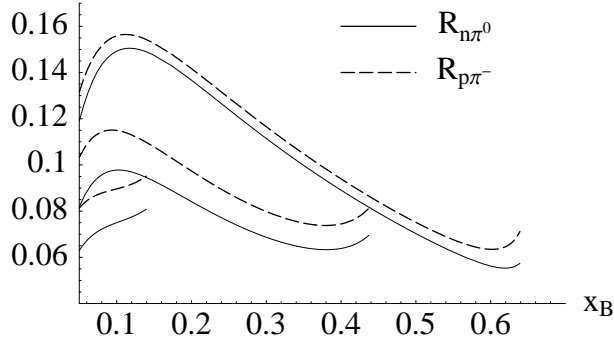}\end{center}

\caption{\label{RNpin_fig} The two ratios of the unpolarized longitudinal
cross sections with and without soft pion emission, $R_{\mathrm{n}\pi^{0}}$
and $R_{\mathrm{p}\pi^{-}}$, as functions of $x_{B}$ for the values
$t_{0}=-1\textrm{ GeV}^{2}$, $t_{0}=-0.3\textrm{ GeV}^{2}$ and $t_{0}=-0.02\textrm{ GeV}^{2}$.
Lower curves belong to smaller values of $-t_{0}$. The photon virtuality
is $Q^{2}=10\textrm{ GeV}^{2}$ and $W_{\mathrm{max}}=1.15\textrm{ GeV}$.
The curves are plotted up to the maximally allowed value of the Bjorken
variable, which is $x_{B\mathrm{max}}=2/(1+\sqrt{1-4M^{2}/t_{0}})$.}\lyxline{\normalsize}

\end{figure}

Figure \ref{RNpin_fig} shows the ratio $R_{N\pi}$ as a function
of $x_{B}$ for three different values of $t_{0}$.%
 We find that for a large range of $t_{0}$, the soft-pion contamination
is roughly 10\% for each individual channel. In situations where we
have to add up both contributions, we arrive at a soft pion contamination
that accounts for 15\% to 30\% within the presented range of $t_{0}$.
The $W_{\mathrm{max}}$-dependence near threshold can easily be deduced
from our results given at $W_{\mathrm{max}}=1.15\textrm{ GeV}^{2}$,
because it is incorporated explicitely in (\ref{RNpin}) through the
phase space function $\Phi(W_{\mathrm{max}})$ (\ref{Phi}).

\subsubsection{Transverse spin asymmetry}

\begin{figure}[!t]
\begin{center}\includegraphics[%
  width=0.50\textwidth]{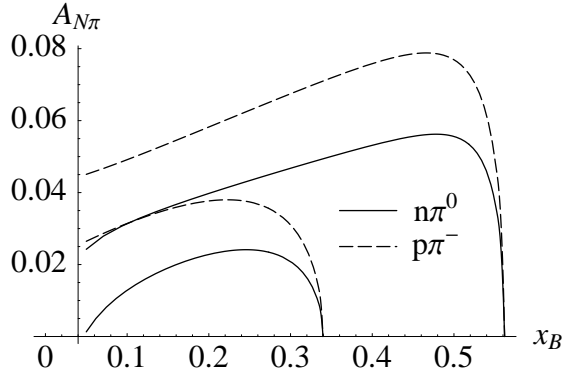}\end{center}

\caption{\label{ANpi_fig} The transverse spin asymmetries of the processes
$p+\gamma^{*}\rightarrow n+\pi_{\mathrm{soft}}^{0}+\pi^{+}$ and $p+\gamma^{*}\rightarrow p+\pi_{\mathrm{soft}}^{-}+\pi^{+}$
for the momentum transfers $t_{0}=-0.3\textrm{ GeV}^{2}$ (lower curves)
and $t_{0}=-1\textrm{ GeV}^{2}$ (upper curves). The photon virtuality
is chosen as $Q^{2}=10\textrm{ GeV}^{2}$. }\lyxline{\normalsize}

\end{figure}
\begin{figure}[!t]
\begin{center}\includegraphics[%
  width=0.52\textwidth]{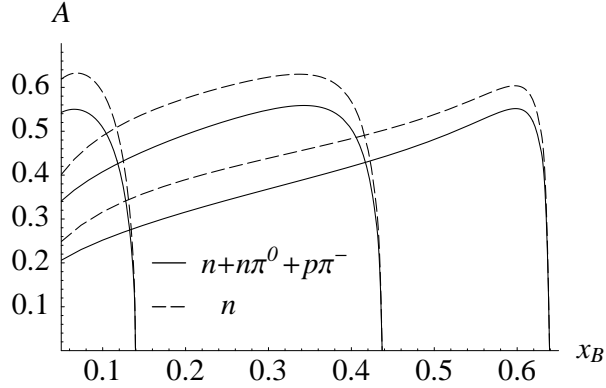}\end{center}

\caption{\label{A_fig} The transverse spin asymmetry for the pure hard $\pi^{+}$
production, $\mathcal{A}_{\mathrm{n}}$ (dashed curve), and for the
hard $\pi^{+}$ production with soft pion admixture, $\mathcal{A}_{\mathrm{n}+\mathrm{n}\pi^{0}+\mathrm{p}\pi^{-}}$
(solid curve). The values of the momentum transfer are $t_{0}=-0.02\textrm{ GeV}^{2}$
(left), $t_{0}=-0.3\textrm{ GeV}^{2}$ (middle) and $t_{0}=-1\textrm{ GeV}^{2}$
(right). In the case of $\mathcal{A}_{\mathrm{n}+\mathrm{n}\pi^{0}+\mathrm{p}\pi^{-}}$,
the maximal invariant mass is chosen to be $W_{\mathrm{max}}=1.15\textrm{ GeV}$.
The photon virtuality is $Q^{2}=10\textrm{ GeV}^{2}$.}\lyxline{\normalsize}

\end{figure}

The results for the transverse spin asymmetries of the processes with
soft pion emission (\ref{HEMP+pi0}) and (\ref{HEMP+pim}) are shown
in figure \ref{ANpi_fig}. The values for these asymmetries turn out
to be very small compared to those of the pure process (\ref{HEMP}),
see figure \ref{A_fig}. The small size of the asymmetries $\mathcal{A}_{N\pi}$
can be traced back to a particularly small value of $s_{1}^{(N\pi)}$
that arises from a significant cancellation within the following terms:
\begin{equation}
s_{1}^{(N\pi)}\propto\textrm{Im}A_{N\pi}\textrm{Re}C_{N\pi}-\textrm{Re}A_{N\pi}\textrm{Im}C_{N\pi}.\end{equation}
In contrast, such a cancellation does not happen in the calculation
of $s_{1}^{(\mathrm{n})}$, because in the pion-pole model for $\tilde{E}$,
(see appendix \ref{GPDs_app}) the imaginary part of $C_{\mathrm{n}}$
vanishes, whereas $A_{\mathrm{n}}$ has non-negligible real and imaginary
parts,\begin{eqnarray}
A_{\mathrm{n}} & = & -\int\limits _{-1}^{1}\d x\,\left(\frac{2/3}{x-\xi+\i0}-\frac{1/3}{x+\xi-\i0}\right)\tilde{H}^{(V)}(x,\xi,\Delta^{2}),\\
C_{\mathrm{n}} & = & \xi\int\limits _{-1}^{1}\d x\left(\frac{2/3}{x-\xi+\i0}-\frac{1/3}{x+\xi-\i0}\right)\tilde{E}^{(V)}(x,\xi,\Delta^{2})=-\frac{3}{2}G_{P}^{(V)}(\Delta^{2}).\end{eqnarray}
Moreover, the values of $s_{0}^{(N\pi)}$ are larger than those of
$s_{0}^{(\mathrm{n})}$, roughly speaking by a factor two or three.
Since they occur in the denominator, $\mathcal{A}_{N\pi}=2s_{1}^{(N\pi)}/(\pi s_{0}^{(N\pi)}),$
this gives a further reduction of the asymmetry $\mathcal{A}_{N\pi}$.

Let us now come to the asymmetry $\mathcal{A}_{\mathrm{n}+\mathrm{n}\pi^{0}+\mathrm{p}\pi^{-}}$
of the combined processes, see equation (\ref{A_n+npi0+ppi-}). Our
discussion of the functions $s_{0}$ and $s_{1}$ above already indicates
that the inclusion of soft pions leads to an certain reduction of
the asymmetry. This result is shown in figure \ref{A_fig}. For an
invariant mass integrated up to $W_{\mathrm{max}}=1.15\textrm{ GeV}$,
we find a downwards shift of the curves of about 10\%, respectively.%

\section{Summary}

In the following, we summarize our statements about soft pion emission
in hard exclusive reactions:

\begin{itemize}
\item We have formulated a \emph{parametrization} of twist-2 lightcone operators
between initial nucleon and final nucleon-pion state. It has turned
out that four functions, which we call $\pi N$ \emph{parton distributions},
are needed for each of these matrix elements. Further, we have shown
how this number is reduced to two when the pion is exactly at the
production threshold.
\item Next, we demonstrated on several examples that the \emph{moments of
the $\pi N$ distributions are polynomials} in the skewedness variables
$\xi$ and $\alpha$. The coefficients in these polynomials are essentially
the form factors (or, following the terminology of electroweak pion
production, the {}``invariant amplitudes'') of pion emission induced
by corresponding \emph{local} twist-2 operators.
\item We have derived \emph{soft-pion theorems for the $\pi N$ distributions}
which represent the leading terms in an expansion in small pion momentum
and mass at given momentum transfer. The nontrivial ingredients of
the final expressions are nucleon and pion GPDs as well as the pion
distribution amplitude. If the momentum transfer is fixed to be large
compared to the pion mass, we find that our results agree with Guichon
et.~al.~\cite{Guichon:2003ah}. In the opposite case, i.e.~when
the momentum transfer is small, we have argued on the example of certain
moments that our results are consistent with the leading order of
ChPT.
\item Moreover, we have provided explicitely a parametrization and the soft-pion
theorems for the resulting form factors of \emph{pion emission from
the nucleon induced by the energy-momentum tensor}.
\item Finally, we have given analytical results for the \emph{amplitude
of hard pion production with soft pion emission} and some numerical
estimates for the particular case of hard $\pi^{+}$ production using
certain GPD models. In the presented kinematical region ($x_{B}\ge0.05$,
$0.02\le-\Delta_{0}^{2}/\textrm{GeV}^{2}\le1$, $W_{\mathrm{max}}=1.15\textrm{ GeV}$,
$Q^{2}=10\textrm{ GeV}^{2}$), we have obtained that the \emph{contamination
of the longitudinal cross section} of the reaction $\gamma^{*}+\textrm{p}\rightarrow\pi^{+}+\textrm{n }$caused
by soft pions amounts to roughly 10\% within each of the two possible
channels $\gamma^{*}+\textrm{p}\rightarrow\pi^{+}+\textrm{n}+\pi_{\mathrm{soft}}^{0}$
and $\gamma^{*}+\textrm{p}\rightarrow\pi^{+}+\textrm{p}+\pi_{\mathrm{soft}}^{-}$.
The effect of the soft-pion channels on the \emph{transverse spin
asymmetry} is a downwards shift of roughly 10\%. Further, we have
observed that the transverse spin asymmetry of the individual soft-pion
reactions is about one order of magnitude smaller than that one which
is predicted for the familiar process without soft pion.
\end{itemize}
We conclude that soft-pion theorems can provide useful estimates of
the soft-pion contamination in hard exclusive reactions. The numerical
studies for the contamination of hard $\pi^{+}$ production indicate
that for an accurate interpretation of experiments an appropriate
separation of the soft-pion channels is required. Second, we point
out that in principle, according to the soft-pion theorems, hard exclusive
reactions with soft pion emission can serve as an additional source
to extract information about GPDs. Therefore, it is worth to consider
corresponding experiments as well as further theoretical studies.

\section*{Acknowledgments}
We acknowledge useful discussions with N.~Kivel, and M.~Vanderhaeghen.
This work is supported in parts
by the Sofja Kovalevskaja Programme of the Alexander von Humboldt
Foundation,  by BMBF and DFG.
\newpage

\part*{Appendix}

\appendix

\section{GPD models}

\label{GPDs_app}

In this appendix, we summarize the models that we have used for the
isovector nucleon GPDs which enter the presented calculations of hard
$\pi^{+}$ production with soft pion emission. They are taken from
reference \cite{Goeke:2001tz}, to which we refer for further explanations
and motivations.

\subsection{Isovector GPD $H^{(V)}$}

The isovector and isoscalar GPDs $H^{(V)}$ and $H^{(S)}$ are linear
combinations of the up and down quark GPDs in the proton,\begin{equation}
H^{(V)}=H^{u}-H^{d},\qquad H^{(S)}=H^{u}+H^{d}.\end{equation}
The models for the quark GPDs $H^{u}$ and $H^{d}$ contain the so-called
double distribution term and the $D$-term. The latter is supposed
to be flavor-independent, hence it cancels in the isovector combination.
Further, one assumes a factorized ansatz for the quark GPDs $H^{q}$,
$q=u,d$, so that\begin{equation}
H^{(V)}(x,\xi,t)=H_{DD}^{u}(x,\xi)\frac{F_{1}^{u/p}(t)}{2}-H_{DD}^{d}(x,\xi)F_{1}^{d/p}(t),\end{equation}
where the double distribution representation is\begin{equation}
H_{DD}^{q}(x,\xi)=\int\limits _{-1}^{1}\d\beta\int\limits _{-(1-|\beta|)}^{1-|\beta|}\d\alpha\,\delta(x-\beta-\alpha\xi)\, h(\beta,\alpha)\, H^{q}(\beta,0,0).\end{equation}
Here, the profile function $h$ is chosen as\begin{equation}
h(\beta,\alpha)=\frac{3}{4}\frac{(1-|\beta|)^{2}-\alpha^{2}}{(1-|\beta|)^{3}},\end{equation}
and $H^{q}(x,0,0)$ denotes the forward limit that is related to the
usual quark distribution functions $q(x)$ and antiquark distribution
functions $\bar{q}(x)$ according to\begin{equation}
H^{q}(x,0,0)=q(x)\,\theta(x)-\bar{q}(-x)\,\theta(-x).\end{equation}
For the numerical calculations, we use the leading order parametrization
MRST2001LO from reference \cite{Martin:2002dr}. The second ingredient
of the factorized ansatz for $H$ is the Dirac form factor $F_{1}^{q/f}$.
It is defined according to\begin{equation}
\langle N_{f}(p')|\bar{\psi}_{q}\gamma_{\mu}\psi_{q}|N_{f}(p)\rangle=\bar{U}'\left[F_{1}^{q/f}\gamma_{\mu}+F_{2}^{q/f}\frac{\i\sigma_{\mu\nu}(p'-p)^{\nu}}{2M}\right]U,\qquad f=p,n,\end{equation}
and related to the nucleon Sachs form factors via\begin{eqnarray}
G_{M}^{(f)}(t) & = & \frac{2}{3}[F_{1}^{u/f}(t)+F_{2}^{u/f}(t)]-\frac{1}{3}[F_{1}^{d/f}(t)+F_{2}^{d/f}(t)]\\
G_{E}^{(f)}(t) & = & \frac{2}{3}\left[F_{1}^{u/f}(t)+\frac{tF_{2}^{u/f}(t)}{4M^{2}}\right]-\frac{1}{3}\left[F_{1}^{d/f}(t)+\frac{tF_{2}^{d/f}(t)}{4M^{2}}\right].\end{eqnarray}
As input for numerical calculations, we have used the empirical fits
of Brash et.~al.~\cite{Brash:2001qq} for the proton Sachs form
factors and those of Bosted \cite{Bosted:1995tm} for the neutron
ones.

\subsection{Isovector GPD $E^{(V)}$}

For $E^{(V)}$, we take the very simple model\begin{equation}
E^{(V)}(x,\xi,t)=[E_{DD}^{u}(x,\xi)-E_{DD}^{d}(x,\xi)]G_{D}(t)\end{equation}
with the double distribution representation\begin{equation}
E_{DD}^{q}(x,\xi)=\int\limits _{-1}^{1}\d\beta\int\limits _{-(1-|\beta|)}^{1-|\beta|}\d\alpha\,\delta(x-\beta-\alpha\xi)\, E^{q}(\beta,0,0)\, h(\beta,\alpha).\end{equation}
 Concerning the forward limit of $E^{q}$, the ansatz\begin{equation}
E^{u}(x,0,0)=\frac{\kappa^{u}}{2}u_{v}(x)\,\theta(x),\qquad E^{d}(x,0,0)=\kappa^{d}d_{v}(x)\,\theta(x)\end{equation}
with\[
\kappa^{u}=2\kappa^{p}+\kappa^{n}\qquad\kappa^{d}=\kappa^{p}+2\kappa^{n}\]
is assumed. Here, $\kappa^{p}=1.793$ and $\kappa^{n}=-1.913$ are
the anomalous magentic moments of the proton and neutron, and $q_{v}(x)$
denotes the valence quark distribution $q(x)-\bar{q}(x)$. In the
$t$-dependent factor of the ansatz for $E^{(V)}$, $G_{D}$ is the
dipole form factor\begin{equation}
G_{D}(t)=\frac{1}{(1-t/0.71\,\mathrm{GeV}^{2})^{2}}.\end{equation}

\subsection{Isovector GPD $\tilde{H}^{(V)}$}

The GPD $\tilde{H}^{(V)}=\tilde{H}^{u}-\tilde{H}^{d}$ is modeled
by the double distribution ansatz\begin{equation}
\tilde{H}^{q}(x,\xi,t)=\tilde{H}_{DD}^{q}(x,\xi)\, F_{A}^{q}(t)\end{equation}
with\begin{equation}
\tilde{H}_{DD}^{q}(x,\xi)=\int\limits _{-1}^{1}\d\beta\int\limits _{-(1-|\beta|)}^{1-|\beta|}\d\alpha\,\delta(x-\beta-\alpha\xi)\, h(\beta,\alpha)\,\Delta q_{v}(\beta),\end{equation}
where $\Delta q_{v}=[\Delta q(x)-\Delta\bar{q}(x)]\theta(x)$ is the
polarized valence quark distribution. For the numerical input of $\Delta q_{v}$
we have used the leading order analysis LSS2001LO given in reference
\cite{Leader:2001kh}. Finally, the axial form factors are approximated
by the dipole form of $F_{A}^{(V)}(t)$,\begin{equation}
F_{A}^{(V)}(t)=\frac{G_{A}^{(V)}(t)}{g_{A}}=\frac{1}{(1-t/M_{A}^{2})^{2}}\end{equation}
with an axial mass $M_{A}\approx1\textrm{ GeV}$.

\subsection{Isovector GPD $\tilde{E}^{(V)}$}

The GPD $\tilde{E}^{(V)}$ is modeled by the pion pole form\begin{equation}
\tilde{E}^{(V)}=G_{P}^{(V)}(t)\frac{1}{\xi}\phi_{\pi}(x/\xi)\,\theta(\xi-|x|),\end{equation}
where the pseudoscalar form factor is assumed to be\begin{equation}
G_{P}^{(V)}(t)=\frac{(2M)^{2}G_{A}^{(V)}(t)}{m_{\pi}^{2}-t}\end{equation}
and for the pion distribution amplitude $\phi_{\pi}$, we use its
asymptotic form $\phi_{\mathrm{as}}$ already given in (\ref{phi_as}).

\section{Soft-pion theorems for non-threshold pion-nucleon distributions}

\label{non-threshold_app}

In section \ref{section_SPT_application}, we have shown the soft-pion
theorems for the threshold $\pi N$ distributions. Here, we give the
corresponding general results, i.e.~when the pion slightly deviates
from the threshold. Throughout the following expressions, we shall
imply the relations\begin{equation}
u=2M^{2}+m_{\pi}^{2}-W^{2}-t+\Delta^{2},\label{u}\end{equation}
\begin{equation}
\bar{\alpha}=\alpha(1-\xi),\end{equation}
\begin{equation}
\xi_{0}=\frac{(p-p')\cdot n}{(p+p')\cdot n}=\frac{2\xi+\bar{\alpha}}{2-\bar{\alpha}}=\frac{2\xi+\bar{\alpha}}{2}[1+\mathcal{O}(\varepsilon)],\end{equation}
and\begin{equation}
\bar{p}_{t}\cdot n=\frac{(p-p'+k)\cdot n}{2}=\xi+\bar{\alpha},\end{equation}
 which make the dependence of the $\pi N$ distributions $H_{i}^{(0,\pm)}$
on the set of variables $x$, $\xi$, $\Delta^{2}$, $\alpha$, $t$,
and $W^{2}$ explicit.

\subsection{Isovector vector $\pi N$ distributions}

\begin{eqnarray}
H_{1}^{(+)} & = & E^{(V)}(x,\xi,\Delta^{2})+\frac{2M^{2}\bar{\alpha}}{u-M^{2}}[H^{(V)}(x,\xi,\Delta^{2})+E^{(V)}(x,\xi,\Delta^{2})]\\
H_{2}^{(+)} & = & 0\\
H_{3}^{(+)} & = & -\left(\frac{M^{2}}{W^{2}-M^{2}}+\frac{M^{2}}{u-M^{2}}\right)\, E^{(V)}(x,\xi,\Delta^{2})\\
H_{4}^{(+)} & = & -\left(\frac{M^{2}}{W^{2}-M^{2}}+\frac{M^{2}}{u-M^{2}}\right)[H^{(V)}(x,\xi,\Delta^{2})+E^{(V)}(x,\xi,\Delta^{2})],\end{eqnarray}
\begin{eqnarray}
H_{1}^{(-)} & = & \frac{\xi_{0}}{g_{A}}[\tilde{E}^{(V)}(x,\xi_{0},t)-\tilde{E}_{\mathrm{pole}}^{(V)}(x,\xi_{0},t)]-\frac{2M^{2}\bar{\alpha}}{u-M^{2}}[H^{(V)}(x,\xi,\Delta^{2})+E^{(V)}(x,\xi,\Delta^{2})]\nonumber \\
 &  & +\frac{4M^{2}}{m_{\pi}^{2}-t}H_{\pi}^{(V)}\left(\frac{x}{\bar{p}_{t}\cdot n},\frac{\xi}{\bar{p}_{t}\cdot n},0\right)\,\theta(\bar{p}_{t}\cdot n-|x|)\\
H_{2}^{(-)} & = & -\frac{\tilde{H}^{(V)}(x,\xi_{0},t)}{g_{A}}+H^{(V)}(x,\xi,\Delta^{2})+E^{(V)}(x,\xi,\Delta^{2})\\
H_{3}^{(-)} & = & -\left(\frac{M^{2}}{W^{2}-M^{2}}-\frac{M^{2}}{u-M^{2}}\right)E^{(V)}(x,\xi,\Delta^{2})\\
H_{4}^{(-)} & = & -\left(\frac{M^{2}}{W^{2}-M^{2}}-\frac{M^{2}}{u-M^{2}}\right)[H^{(V)}(x,\xi,\Delta^{2})+E^{(V)}(x,\xi,\Delta^{2})].\end{eqnarray}

\subsection{Isoscalar vector and gluon $\pi N$ distributions}

\begin{eqnarray}
H_{1}^{(0,G)} & = & E^{(S,G)}(x,\xi,\Delta^{2})+\frac{2M^{2}\bar{\alpha}}{u-M^{2}}[H^{(S,G)}(x,\xi,\Delta^{2})+E^{(S,G)}(x,\xi,\Delta^{2})]\nonumber \\
 &  & +\frac{2M^{2}}{m_{\pi}^{2}-t}H_{\pi}^{(S,G)}\left(\frac{x}{\bar{p}_{t}\cdot n},\frac{\xi}{\bar{p}_{t}\cdot n},0\right)\,\theta(\bar{p}_{t}\cdot n-x)\\
H_{2}^{(0,G)} & = & 0\\
H_{3}^{(0,G)} & = & -\left(\frac{M^{2}}{W^{2}-M^{2}}+\frac{M^{2}}{u-M^{2}}\right)\, E^{(S,G)}(x,\xi,\Delta^{2})\\
H_{4}^{(0,G)} & = & -\left(\frac{M^{2}}{W^{2}-M^{2}}+\frac{M^{2}}{u-M^{2}}\right)[H^{(0,G)}(x,\xi,\Delta^{2})+E^{(0,G)}(x,\xi,\Delta^{2})]\end{eqnarray}

\subsection{Axial vector $\pi N$ distributions}

\begin{eqnarray}
\tilde{H}_{1}^{(0,+)} & = & \xi\tilde{E}^{(S,V)}(x,\xi,\Delta^{2})-\frac{2M^{2}\bar{\alpha}}{u-M^{2}}\tilde{H}^{(S,V)}(x,\xi,\Delta^{2})\\
\tilde{H}_{2}^{(0,+)} & = & 0\\
\tilde{H}_{3}^{(0,+)} & = & -\left(\frac{M^{2}}{W^{2}-M^{2}}-\frac{M^{2}}{u-M^{2}}\right)\xi\tilde{E}^{(S,V)}(x,\xi,\Delta^{2})\\
\tilde{H}_{4}^{(0,+)} & = & -\left(\frac{M^{2}}{W^{2}-M^{2}}-\frac{M^{2}}{u-M^{2}}\right)\tilde{H}^{(S,V)}(x,\xi,\Delta^{2}),\end{eqnarray}
\begin{eqnarray}
\tilde{H}_{1}^{(-)} & = & \frac{E^{(V)}(x,\xi_{0},t)}{g_{A}}+\frac{2M^{2}\bar{\alpha}}{u-M^{2}}\tilde{H}^{(V)}(x,\xi,\Delta^{2})\\
\tilde{H}_{2}^{(-)} & = & -\frac{E^{(V)}(x,\xi_{0},t)+H^{(V)}(x,\xi_{0},t)}{g_{A}}+\tilde{H}^{(V)}(x,\xi,\Delta^{2})\\
\tilde{H}_{3}^{(-)} & = & -\left(\frac{1}{2}+\frac{M^{2}}{W^{2}-M^{2}}+\frac{M^{2}}{u-M^{2}}\right)\xi\tilde{E}^{(V)}(x,\xi,\Delta^{2})\nonumber \\
 &  & +\frac{2M^{2}}{g_{A}(m_{\pi}^{2}-\Delta^{2})}\,\phi_{\pi}\left(\frac{x}{\xi}\right)\,\theta(\xi-|x|)\\
\tilde{H}_{4}^{(-)} & = & -\left(\frac{M^{2}}{W^{2}-M^{2}}+\frac{M^{2}}{u-M^{2}}\right)\,\tilde{H}(x,\xi,\Delta^{2}).\end{eqnarray}

\section{Soft-pion theorems for the form factors of pion emission induced
by the local vector current}

\begin{lyxlist}{00.00.0000}
\item [\label{appendix_local_vector}]~
\end{lyxlist}
In equation (\ref{electroproduction_amplitude}) of section \ref{section_moments}
in the main text, we have parametrized the matrix element for pion
emission induced by the local vector current in terms of form factors
$A_{i}$. As explained in this section, the soft-pion theorems for
these form factors can be obtained from the first moments of the $\pi N$
distributions functions $H_{i}^{(0,\pm)}$ that have been given in
appendix \ref{non-threshold_app}. In this way, we obtain for the
isoscalar and isovector even form factors: \begin{eqnarray}
\lefteqn{\sum_{i=1}^{8}U'\, A_{i}^{(0,+)}\Gamma_{i}^{\mu}U}\nonumber \\
 & = & \bar{U}'\left\{ F_{1}^{(S,V)}(\Delta^{2})\,\bar{p}^{\mu}+\frac{2M^{2}[F_{1}^{(S,V)}(\Delta^{2})+F_{2}^{(S,V)}(\Delta^{2})}{u-M^{2}}k^{\mu}\right.\nonumber \\
 &  & -\left(\frac{M}{W^{2}-M^{2}}+\frac{M}{u-M^{2}}\right)F_{2}^{(S,V)}(\Delta^{2})\,\hat{k}\bar{p}^{\mu}\nonumber \\
 &  & \left.-\left(\frac{M^{2}}{W^{2}-M^{2}}+\frac{M^{2}}{u-M^{2}}\right)[F_{1}^{(S,V)}(\Delta^{2})+F_{2}^{(S,V)}(\Delta^{2})]\hat{k}\gamma^{\mu}\right\} \gamma_{5}U+\mathcal{O}(\varepsilon).\end{eqnarray}
To the given accuracy, current conservation is fulfilled:\begin{equation}
\Delta^{\mu}\sum_{i}\bar{U}'A_{i}^{(0,+)}\Gamma_{i}^{\mu}U=\mathcal{O}(\varepsilon).\end{equation}
For the isovector odd form factors, we obtain\begin{eqnarray}
\lefteqn{\sum_{i=1}^{8}U'\, A_{i}^{(-)}\Gamma_{i}^{\mu}U}\nonumber \\
 & = & \bar{U}'\left\{ -\frac{G_{P}^{(V)}(t)}{2g_{A}}\Delta^{\mu}+\left[-\frac{2M^{2}[F_{1}^{(V)}(\Delta^{2})+F_{2}^{(V)}(\Delta^{2})}{u-M^{2}}+\frac{4M^{2}}{m_{\pi}^{2}-t}\right]k^{\mu}\right.\nonumber \\
 &  & +[-F_{A}^{(V)}(t)+F_{1}^{(V)}(\Delta^{2})+F_{2}^{(V)}(\Delta^{2})]M\gamma_{\mu}-\left(\frac{M}{W^{2}-M^{2}}-\frac{M}{u-M^{2}}\right)F_{2}^{(V)}(\Delta^{2})\,\hat{k}\bar{p}^{\mu}\nonumber \\
 &  & -\left(\frac{M}{W^{2}-M^{2}}-\frac{M}{u-M^{2}}\right)F_{2}^{(V)}(\Delta^{2})\,\hat{k}\bar{p}^{\mu}\nonumber \\
 &  & \left.-\left(\frac{M^{2}}{W^{2}-M^{2}}-\frac{M^{2}}{u-M^{2}}\right)[F_{1}^{(V)}(\Delta^{2})+F_{2}^{(V)}(\Delta^{2})]\hat{k}\gamma^{\mu}\right\} \gamma_{5}U+\mathcal{O}(\varepsilon).\end{eqnarray}
Current conservation at small momentum transfer is immediately fulfilled,
but for the moderate momentum transfer, the pseudoscalar form factor
$G_{P}$ cannot simply be approximated by its pion pole form, and
so we arrive at\begin{equation}
\Delta_{\mu}\sum_{i}\bar{U}'A_{i}^{(-)}\Gamma_{i}^{\mu}U=-\frac{M}{g_{A}}\left[G_{A}(t)\,2M+G_{P}(t)\frac{t}{2M}\right]\bar{U}'\gamma_{5}U+\mathcal{O}(\varepsilon)\qquad(-t\gg\varepsilon^{2}).\end{equation}
But since the remaining combination of axial and pseudoscalar form
factor is proportional to $m_{\pi}^{2}$,\begin{equation}
\i\bar{U}'\left[G_{A}(t)\,2M+G_{P}(t)\frac{t}{2M}\right]\gamma_{5}\frac{\tau^{a}}{2}U=\langle N(p')|\partial\cdot A^{a}|N(p)\rangle=f_{\pi}m_{\pi}^{2}\langle N(p')|\Phi^{a}|N(p)\rangle\end{equation}
and pion-pole enhancement at such large $t$ is excluded, it is reasonable
to assume $G_{A}(t)\,2M+G_{P}(t)\, t/(2M)=\mathcal{O}(\varepsilon)$.
Hence for any momentum transfer in the considered region $-t\le M^{2}$
we arrive at \begin{equation}
\Delta^{\mu}\sum_{i}\bar{U}'A_{i}^{(-)}\Gamma_{i}^{\mu}U=\mathcal{O}(\varepsilon),\end{equation}
so that finally, current conservation is fulfilled.


\end{document}